\documentclass[11pt,a4paper]{article}
\usepackage{jheppub}
\usepackage{amsmath,amssymb,mathtools}
\usepackage[normalem]{ulem}

\newcommand{\AdS}{\mathrm{AdS}}

\newcommand{\be}{\begin{equation}}
\newcommand{\ee}{\end{equation}}

\usepackage{soul}

\begin{document}
\title{Dirichlet walls and the end of time}
\author{Hanako~Helton,}
\author{Gary~T.~Horowitz,}
\author{and Donald~Marolf}

\affiliation{Department of Physics, University of California, Santa Barbara, CA 93106, U.S.A.}

\abstract{We study evolution in Einstein-Hilbert gravity with Dirichlet boundary conditions imposed on a finite surface. 
We argue that there are open sets of initial data where such evolutions terminate at finite times due to singularities that reach the boundary. In any dimension, the simplest such examples occur in cosmologies. However, in 2+1 dimensions we also show that Dirichlet walls initially outside a BTZ black hole can fall through the horizon, and that this also leads to generic singularities. A similar construction in higher dimensions leads to trapped surfaces that reach the wall, though the end result of such evolutions is more difficult to study.}

\maketitle

\section{Introduction}

It is well-known that spatially closed universes are generically singular in general relativity. The most general theorem proves the existence of an incomplete causal geodesic \cite{Hawking:1970zqf}, but under somewhat stronger assumptions, one can show the existence of 
spacelike big-bang or big-crunch singularities which forbid further Einstein-Hilbert time evolution anywhere in the entire universe \cite{Hawking:1966jv}. Whether such a picture will be altered by quantum effects or other new physics has been a long-running topic of discussion and debate but, at least at the classical level, singularities of this sort can be described as signaling the end of time.

On the other hand, with familiar four-dimensional asymptotically flat or asymptotically anti-de Sitter (AdS) boundary conditions, spacelike singularities appear to be hidden inside event horizons and, in particular, generally do not reach the spacetime boundary. Exceptions are known only when the boundary conditions become singular as well and, in the AdS case, in particular when any dual field theory would also find itself driven to a singularity \cite{Hertog:2005hu}. While visible naked singularities can also arise from e.g. either Gregory-Laflamme-like instabilities \cite{GregoryLaflamme1993,EmparanReall2006,LehnerPretorius2010} in higher dimensions or from finely-tuned initial data such as that associated with critical collapse \cite{Choptuik1993}, and while they can cut off evolution at infinity, (see \cite{Marolf2010} for discussion of the Gregory-Laflamme context), in both cases quantum effects plausibly allow evolution through such singularities resulting only in the emission of a few Planck-scale particles that then propagate to infinity without significant further consequences. In particular, in the AdS context there is no reason to associate these classical naked singularities in the bulk with singular behavior of any dual field theory.

In contrast, our work below examines the behavior of Einstein-Hilbert gravity with Dirichlet boundary conditions that fix a (finite) Lorentz-signature induced metric on a timelike boundary. We refer to a boundary with such a fixed induced metric as a {\it Dirichlet wall}. Such boundaries have seen much interest in the past few years, in part motivated by so-called $T\bar T$ deformations of AdS/CFT \cite{mcgoughMovingCFTBulk2018}, and they have recently been shown to admit a well-defined initial value problem in certain contexts \cite{anWellposedGeometricBoundary2025a}, though important questions remain concerning the stability of such theories and their long-term evolution \cite{Andrade:2015qea}. 

Our focus will be on the case where the wall metric is static, and more specifically on cases where this metric is just $S^{d-2} \times {\mathbb R}$ for a bulk spacetime of dimension $d$, so that the framework admits a natural notion of conserved energy. Many discussions of Dirichlet walls follow \cite{York1986,mcgoughMovingCFTBulk2018} in studying solutions of this kind and, in particular, in focusing on those constructed by cutting off a standard (e.g., AdS-Schwarzschild) spherically symmetric black hole at some radius $R$. Such solutions may be said to describe black holes in spherical boxes (or {\it boxed black holes}, for short). In these well-known cases, the spacelike singularity is again confined within an event horizon. Furthermore, at least for uncharged black holes, such solutions are known to be perturbatively stable.

Nevertheless, for $S^{d-2} \times {\mathbb R}$ Dirichlet boundary conditions, we will argue below that there are open sets of initial data on which time evolution leads to the formation of spacelike singularities at the location of the Dirichlet wall. Using the terminology of the cosmic censorship literature, such singularities can then be said to be {\it generic} {\cite{Wald:1997wa}}. {These are not curvature singularities, but rather singularities in the sense that it becomes impossible to continue to consistently impose the Dirichlet boundary conditions. These singularities resemble {certain examples of} } the big-bang/big-crunch cosmological singularities mentioned above {e.g., in cosmological spacetimes given by quotients of anti-de Sitter space}. 
Our results thus raise interesting questions about possible quantum effects or other new physics, and in particular about their description in any dual field theory. However, at the classical Einstein-Hilbert level these singularities again appear to signal an end of time.

We begin our analysis of explicitly-singular bulk metrics in section \ref{sec:examples} by investigating a particular class of solutions that one may describe as cosmologies with Dirichlet walls. The initial boundary value problem is again well-posed in this setting by the argument of \cite{anWellposedGeometricBoundary2025a}\footnote{\label{foot:AA} While the Brown-York stress tensor in this case differs by an overall sign from those considered in \cite{anWellposedGeometricBoundary2025a}, this sign does not affect the proof of well-posedness. We thank Michael Anderson for correspondence on this point.}. Such solutions thus fall under the general framework described in \cite{Philcox:2025faf}, which we will call {\it Dirichlet cosmologies}. However, we are interested here in vacuum spacetimes with vanishing or negative cosmological constant $\Lambda$. The focus on $\Lambda\le 0$ is motivated both by potential connections to $T\bar T$ deformations of AdS/CFT and by our desire to discuss big-bang/big-crunch singularities without adding explicit matter fields. We show that a subclass of these solutions exhibit spacelike end-of-time singularities that, in particular, reach the Dirichlet wall. It will then follow by continuity that this must remain true for an open set of initial data around any such solution, {where the open set is taken in the class of quotients of AdS.\footnote{{On physical grounds, introducing {gravitationally-attractive classical} matter {e.g., satisfying the null energy condition,} should not avoid the formation of a singularity. {We thus expect that the } class of solutions within which we may consider open sets of initial data is much larger than just quotients of AdS.}}}

However, such Dirichlet cosmologies are rather different from the more familiar boxed black hole solutions. Indeed, we will see that the particular solutions described in section \ref{sec:examples} can be distinguished from standard boxed black holes by the sign of their conserved energy. This then raises the question of whether the two classes of solutions might perhaps live in separate connected components of the space of classical solutions and, if so, whether one might choose to regard each component as its own separate theory. This would then leave open the possibility that the component associated with familiar boxed black holes might lack the generic end-of-time singularities of our Dirichlet cosmologies.

We therefore address this issue in section 
\ref{sec:curvature} by exhibiting a context in which one can deform continuously from solutions with $E<0$ to those with $E>0$. In particular, while $E$ is of course conserved with time-independent boundary conditions, it is generally possible to drive the system between settings with different signs of $E$ by taking the boundary conditions to be time-dependent for limited periods of time. While such changes in the sign of $E$ necessarily entail violations of the conditions under which well-posedness was proven in \cite{anWellposedGeometricBoundary2025a}, we argue that physics requires us to allow such transgressions as they occur under very general conditions when gravitational radiation exerts significant back-reaction on the background spacetime near the wall. {We also note that, if the theory ceases to be well-posed at some point along our deformation, this is again a kind of singular behavior that has arises at a finite point.}

We then argue in section \ref{sec:WallinBTZ} that time-independent Dirichlet walls can encounter singularities even when $E$ has the same sign as for boxed black holes. To do this, we specialize to the case of 2+1 vacuum solutions and consider initial data that describes a Dirichlet wall surrounding a Ba\~nados-Teitelboim-Zanelli (BTZ) black hole. In that case, if {even a small piece of} a Dirichlet wall falls through the black hole horizon, {the entire wall} will necessarily encounter a singularity in finite time and, in fact, at least locally this singularity is much like the end-of-time singularities of our Dirichlet cosmologies.\footnote{The first piece of the wall that falls through the horizon turns out to have {positive} Brown-York energy density, matching the sign for which section \ref{sec:examples} found end-of-time singularities.} We will construct walls where this occurs. 

The tentative extension to higher dimensions of this line of reasoning is then discussed in section \ref{sec:trapped}. In that context it is straightforward to argue that trapped surfaces form that touch the Dirichlet wall. However, it is then more complicated to extract results about the formation of singularities at later times. We close with some final discussion in section \ref{sec:disc} where we comment on generalizations to other boundary conditions, and future directions.

\section{Singularities in Dirichlet cosmologies}
\label{sec:examples}

The fact that Dirichlet walls are finite surfaces in spacetime leads to a critical difference from both the asymptotically flat and asymptotically AdS contexts. In particular, when there are two Dirichlet walls, they can collide with each other at finite times. This section will study several examples of singularities analogous to such collisions with just a single Dirichlet wall. We will then argue that they can form on open sets of initial data.

We begin in section \ref{subsec:flatcol} by briefly discussing a local model of such collisions given by flat Dirichlet walls in empty Minkowski space. We then argue for similar behavior in general AdS$_d$ cosmologies in section \ref{sec:genarg} for $d\ge 3$. Analogous results for vanishing cosmological constant also follow by taking the usual limit. Sections \ref{sec:ads3compactifications} and \ref{sec:conicalsingularities} then analyze the AdS$_3$ case in more detail, while section \ref{sec:curvature} comments on settings with positive cosmological constant.

\subsection{When walls collide}
\label{subsec:flatcol}

\begin{figure} [h!]
    \centering
    \includegraphics[width=0.3\linewidth]{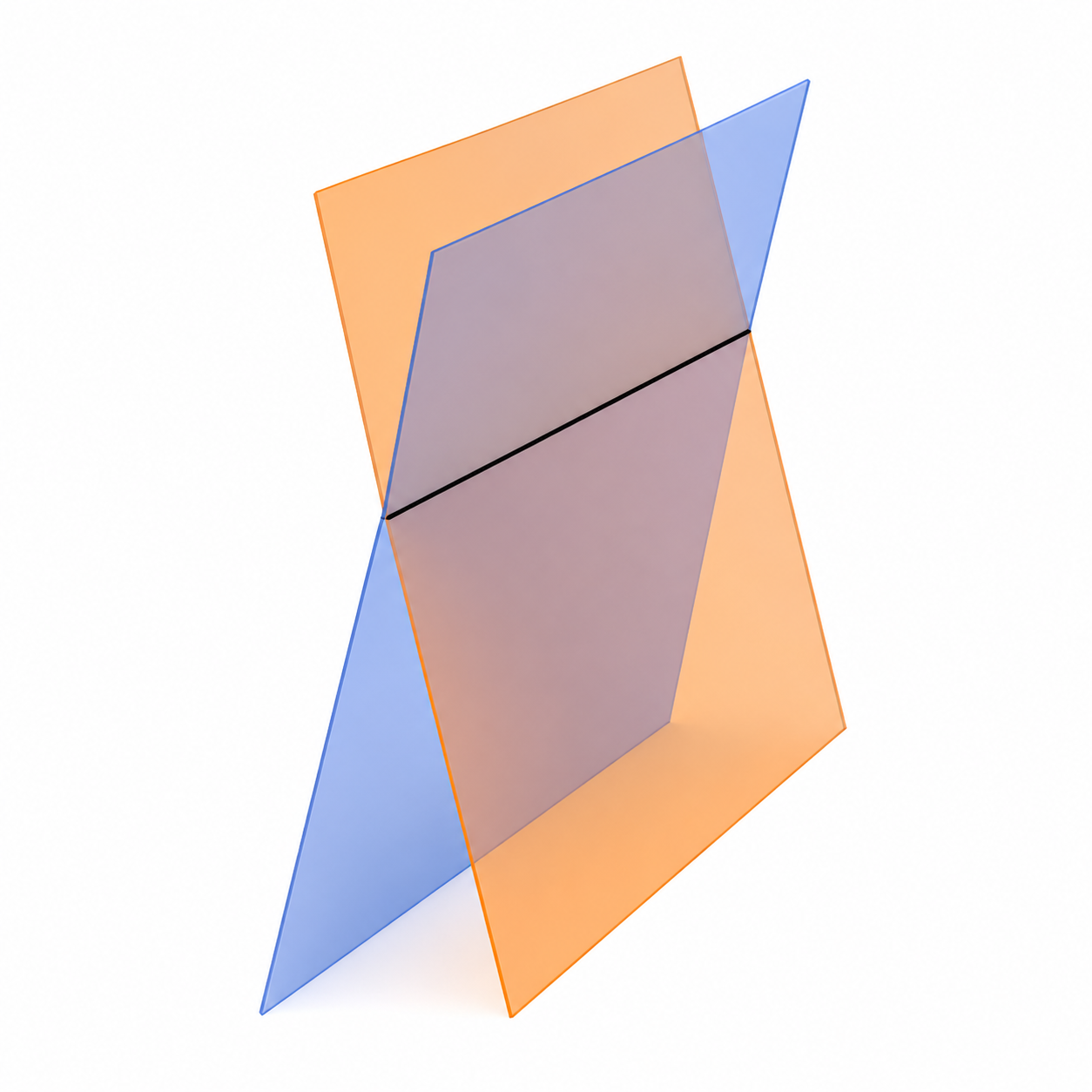}
    \includegraphics[width=0.3\linewidth]{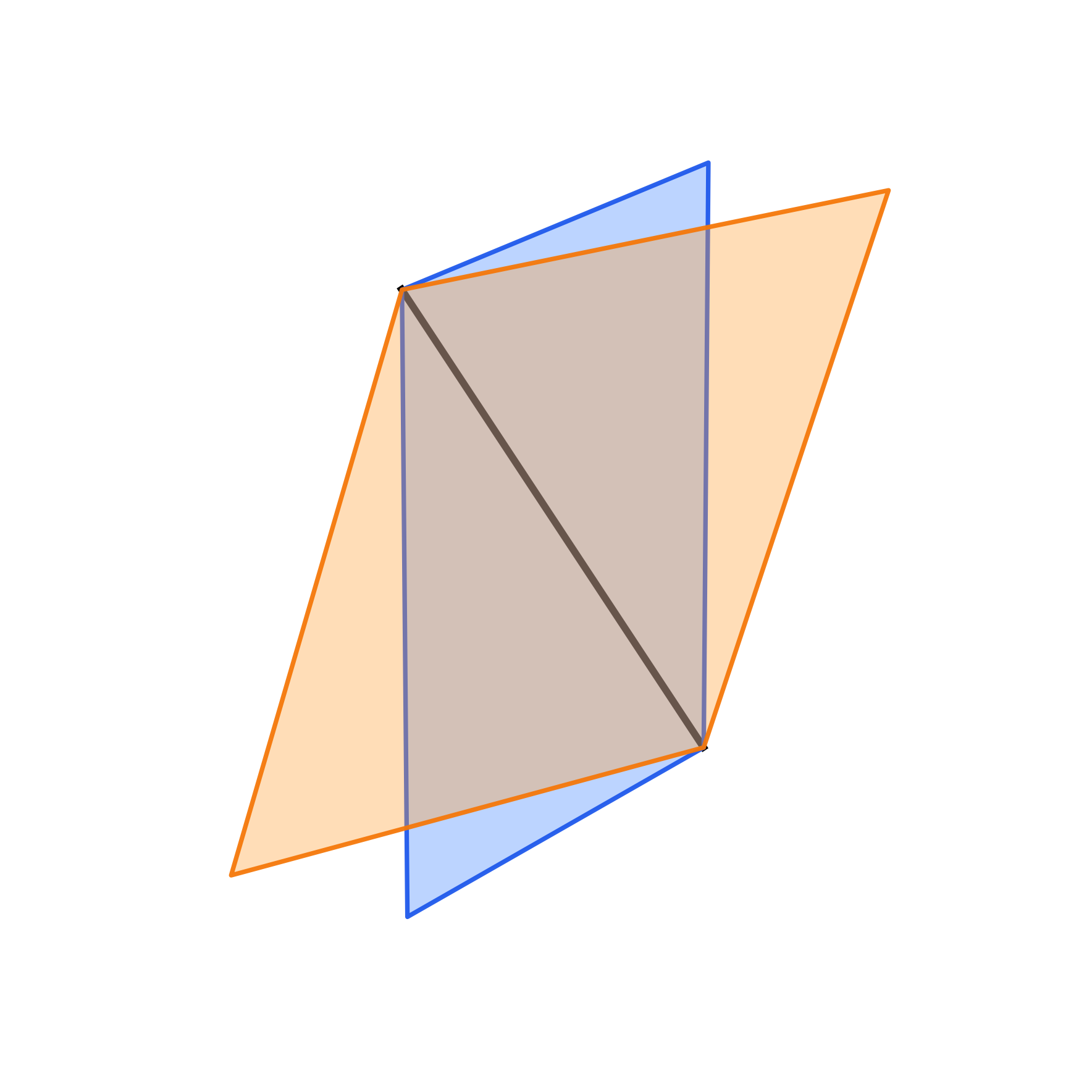}
    \includegraphics[width=0.3\linewidth]{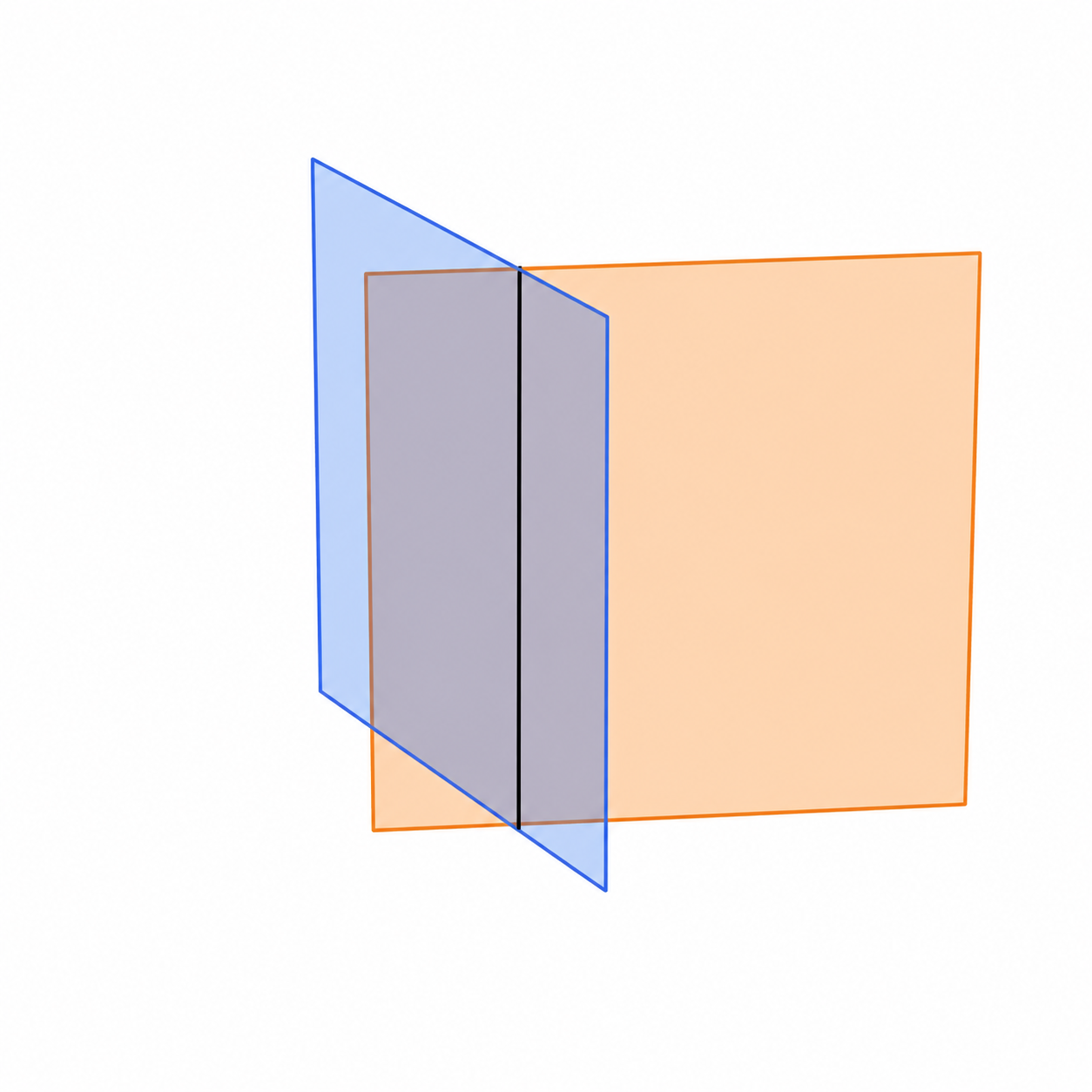}
    \caption{Unless they are precisely parallel, two timelike planes in Minkowski intersect along a straight line. The intersection is spacelike when the orientations differ by a boost (left), it is null when they differ by a null rotation (center), and it is timelike when they differ by a rotation (right).}
    \label{fig:tplanes}
\end{figure}

Since any smooth spacetime is locally Minkowski, and since any smooth wall is locally flat, 
a local model of a generic collision of Dirichlet walls should be obtained by studying the spacetime between two planar Dirichlet walls in flat Minkowski space. If the walls are parallel, there are no singularities and the solution exists for all time. If the walls are parallel in space, but one is boosted with respect to the other, then they will intersect in a spacelike surface. This completely cuts off the entire spacetime between the walls; see Figure \ref{fig:tplanes} (left). If there is no relative boost, but one wall is rotated relative to the other, then they intersect in a timelike surface as in Figure \ref{fig:tplanes} (right). In both cases, the intersection is a singularity in the sense that one cannot maintain the flat metric on the wall past the intersection. 

For spacelike intersections we have thus constructed an end-of-time singularity. Any such singularity is clearly generic, as a small perturbation of a spacelike intersection must again be spacelike. Furthermore, since the above spacetimes then have translation symmetries along the codimension-2 spacelike plane on which our walls intersect, we can remove unwanted asymptotic regions by periodically compactifying our spacetime in such directions. 

Of course the spacetimes with end-of-time singularities described above have Dirichlet walls that are very different from those in the boxed black hole spacetimes. In particular, the walls used above have the property that they have two distinct connected components on spacelike initial data surfaces. In contrast, the boundaries typically used to box in black holes are fully connected on spacelike initial surfaces. However, the various subsections below will showcase examples where this distinction makes little difference to the dynamics. In particular, for the rest of section \ref{sec:examples} we will confine ourselves to the study of static walls with induced metric $S^{d-2} \times {\mathbb R}.$ In order to make greater contact with potential $T\bar T$-like deformations of AdS/CFT, we will also focus on the case of negative cosmological constant $\Lambda$, though it will be straightforward to take the limit $\Lambda \rightarrow 0$ if one desires. Some comments on the case $\Lambda>0$ will appear in section \ref{sec:curvature}. 

Before proceeding to the case of $S^{d-2}\times {\mathbb R}$ boundaries, there is one more lesson to be learned from our local model. As we have seen, for general orientations of the two walls discussed above, the intersection will be a flat codimension-2 surface which can be spacelike, timelike, or null. But unless it is spacelike, the initial data on any compact spacelike surface will be singular at all times. Since we will only consider nonsingular initial data on compact spacelike surfaces, the singularities we find will always start off spacelike.

\subsection{$S^{d-2} \times {\mathbb R}$ Dirichlet walls and singularities for general $d$}
\label{sec:genarg}

Compactifications of anti-de Sitter space provide a natural arena in which to study evolution with a Dirichlet wall. $\AdS_d$ has a symmetry group $SO(d-1, 2)$. Quotienting by a discrete group $\Gamma$ can yield a spatially closed manifold. As is well known, the resulting spacetimes contain both initial and final singularities. We will show below that such singularities will still form when we insert boundaries with fixed induced metric $S^{d-2} \times {\mathbb R}$. 
In particular, the singularities will then reach the Dirichlet wall. In the present section we give a general argument that applies in all dimensions. Further details (and some generalizations) for the case of $\AdS_3$ will then be given in section~\ref{sec:ads3compactifications}.

To begin, recall that anti-de Sitter space can be written as the (universal cover of the) hyperboloid defined by
\begin{equation}
    - T_1^2 - T_2^2 + X_1^2 + \dots + X_{d-1}^2 = - \ell^2,
\end{equation}
where in the embedding coordinates $T_1, T_2, X_i$ the metric is $ds^2 = -d T_1^2 - dT_2^2 + \sum_{i = 1}^{d-1}dX_i^2.$ $\AdS_d$ has a symmetry group $SO(d-1, 2)$ inherited from the embedding space $M^{2,d-1}$ (Minkowski space with signature ($2,d-1$)). The $\AdS_d$ metric can be written in the {comoving} coordinates defined by 
\begin{equation}
\label{eq:sigmachi}
    T_1 = \sin \tau, \quad T_2 = \cos \tau \cosh \chi, \quad X_i = \cos \tau \sinh \chi~\hat{x}_i,
\end{equation}
where $\hat{x}_i$ satisfy $\hat{x}_i \hat{x}^i = 1$ and hence define a unit $(d-2)$-sphere, and where we have set the AdS radius $\ell$ to unity. Here, a constant $\tau$ surface is a constant negative curvature hyperboloid of radius $\cos \tau$. The metric becomes
\begin{equation} \label{eq:metric}
    ds^2 = - d\tau^2 + \cos^2 \tau \, d\sigma_{d-1}^2.
\end{equation}
where $d\sigma_{d-1}^2 = d\chi^2 + \sinh^2{\chi}\ d\Omega_{d-2}^2$ is the metric on the unit hyperbola. Using these coordinates puts the metric in Friedman-Robertson-Walker form, which will be convenient for cosmological solutions. The coordinate system covers the domain of dependence of the $\tau = 0$ slice.

In order to construct the solutions of interest, we quotient anti-de Sitter spacetime by a discrete group $\Gamma$. It is easiest to view these quotients by restricting to a slice of constant $\tau$. Since the above coordinates are comoving, the isometries of the spatial slice are also isometries of the entire spacetime. In particular, the symmetries of the spatial hyperboloid form a $SO(d-1, 1)$ subgroup of the full symmetry group $SO(d-1, 2)$ of $\AdS_d$. The action of the $SO(d - 1, 1)$ subgroup on the spacetime is implied by its action on the $\tau = 0$ surface; this parametrization is convenient for a spacetime with a moment of time-reflection symmetry. 

We choose $\Gamma$ so that the spatial hyperboloids become compact. As usual, instead of taking a quotient, we can alternatively describe each constant-$\tau$ slice of our spacetime as being given by an appropriate fundamental domain within the spatial hyperboloid with the understanding that the edges of this domain are then identified in an appropriate way. Note that the resulting spatially-compact spacetime has singularities at $\tau = \pm \pi/2$ where $\cos \tau$ vanishes and the volume of space shrinks to zero.

\begin{figure}[h!]
    \centering
    \includegraphics[width=0.25\linewidth]{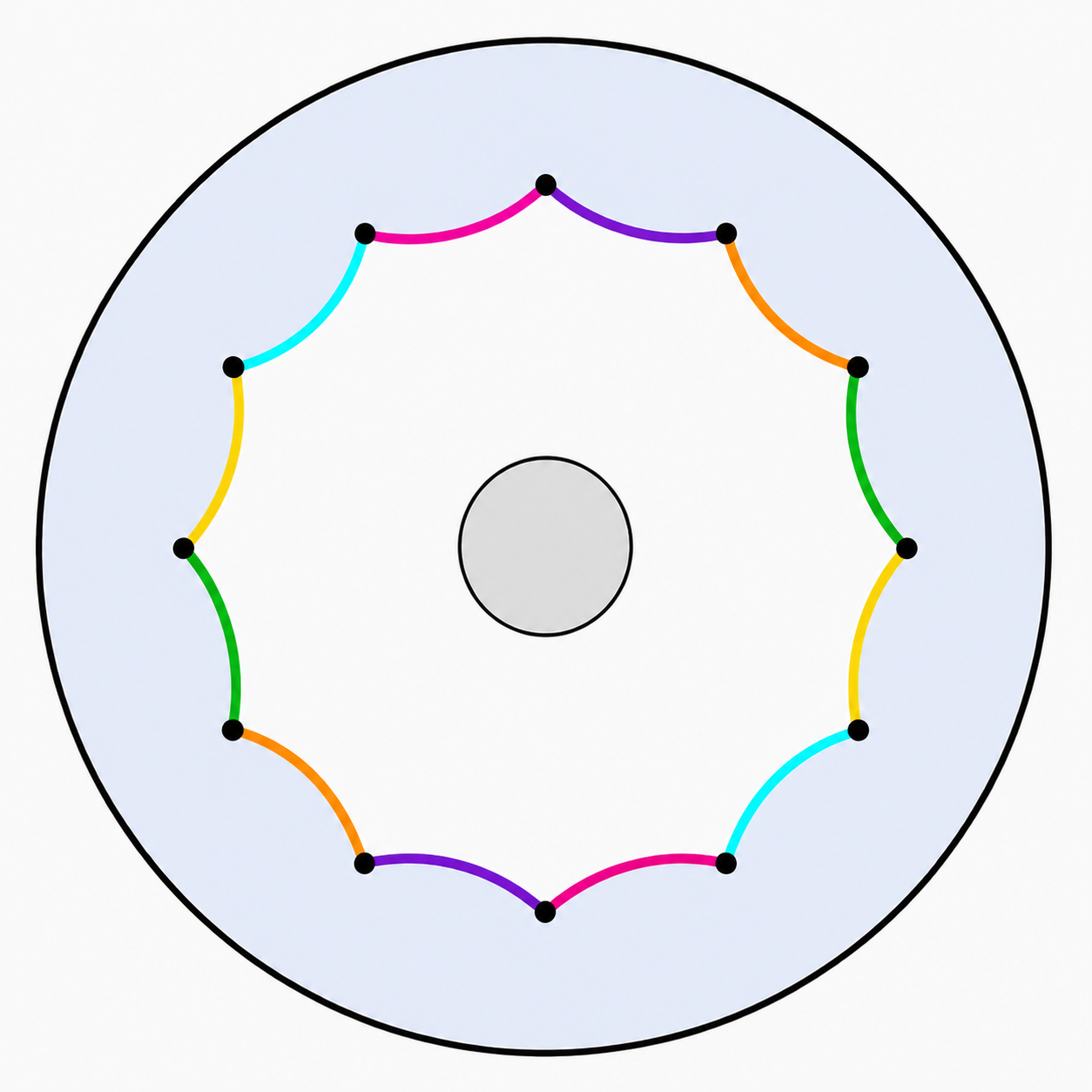}
    \includegraphics[width=0.25\linewidth]{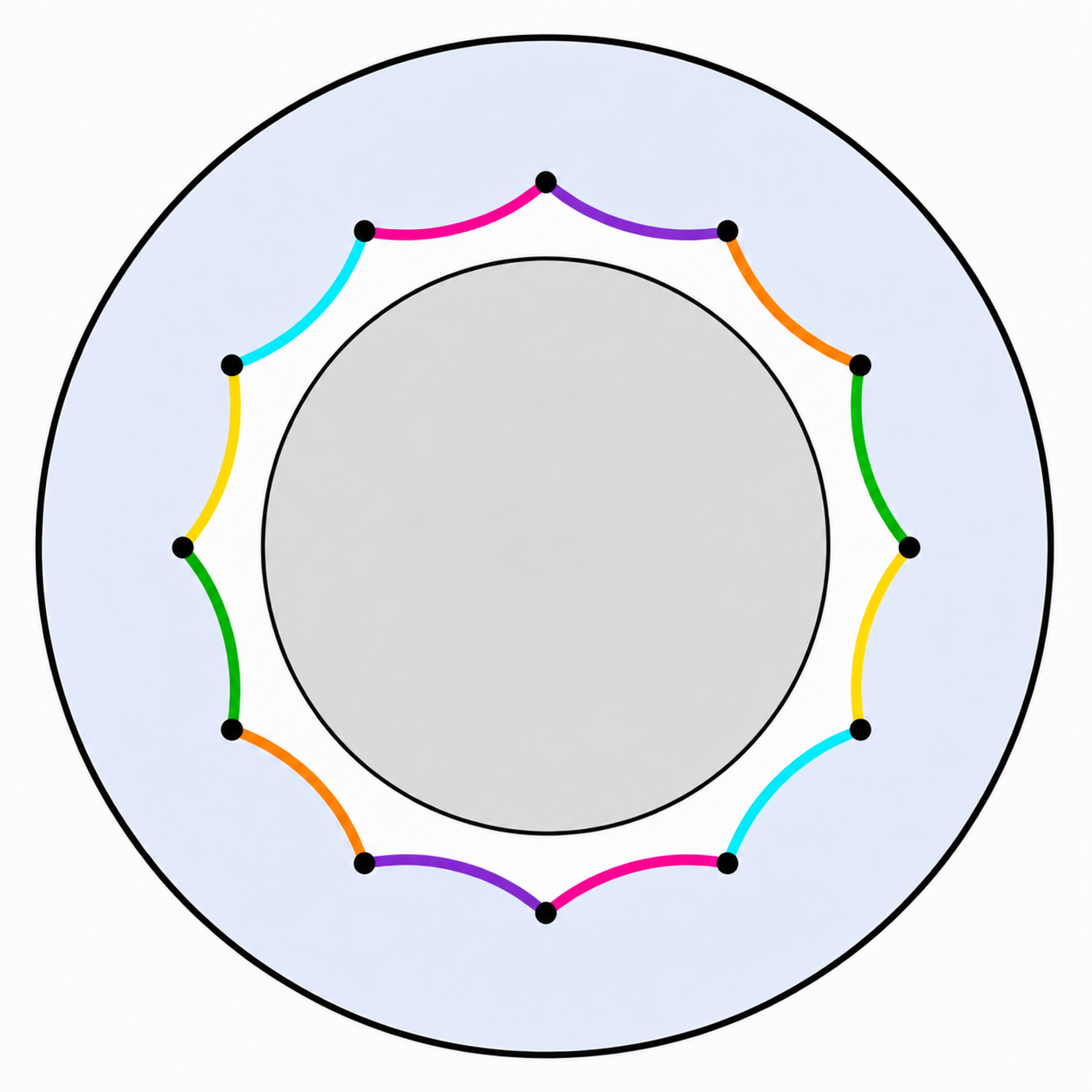}
    \caption{A fundamental region in the hyperbolic plane and possible round Dirichlet walls for the case $d=3$ with spatial slices of genus 3. On the boundary of the fundamental domain, line segments of the same color are identified. To avoid singularities at $\tau = 0$, the Dirichlet wall must fit inside the fundamental domain. The left panel shows a small Dirichlet wall, while the right panel shows a wall of nearly maximal size. As $\tau$ increases from $\tau=0$, the proper size of the fundamental region shrinks while that of a static wall remains constant. Singularities form when the two collide.}
    \label{fig:smallbig1}
\end{figure}

We now introduce the Dirichlet wall. The simplest such wall with induced metric $S^{d-2} \times {\mathbb R}$ is given by a static, spherically symmetric wall, centered at the origin ($\chi=0$). 
The wall is most easily described in global coordinates, in which the AdS metric is
\begin{equation}\label{eq:global}
    ds^2 = - f(r) dt^2 + f(r)^{-1}dr^2 + r^2 \,d \Omega_{d-2}^2,
\end{equation}
where $f(r) = r^2+1.$ We take our boundary to coincide with a constant $r=r_0$ surface. This is consistent with our compactification of the spatial hyperboloid at $\tau = 0$ provided that $r_0$ is small enough to fit inside a fundamental domain of the compactification; see Figure \ref{fig:smallbig1}. 

We consider only the spacetime region {\it outside} this wall (i.e., corresponding to points in the fundamental domain with $r>r_0$). Since \eqref{eq:metric} requires any fundamental domain to contract to a single point in AdS$_d$ at $\tau = \pi/2$, there must be some $\tau< \pi/2$ when the boundary of the fundamental domain collides with our Dirichlet wall. 
Indeed, there must also be some $\tau < \pi/2$ where there are generically-distinct points $x$ and $y$ on our Dirichlet wall that are identified by the action of $\Gamma$. These are points where the Dirichlet wall collides with itself, and thus where the local physics is modeled by the colliding planes of section \ref{subsec:flatcol}. As noted there, since the constant $\tau$ surfaces are spacelike and compact, and since there is no such singularity for $\tau=0$, the resulting wall-collision singularities must be spacelike at the value of $\tau$ where they first form.

However, the resulting physical implications may depend on the details of the further evolution. 
In particular, 
singularities are often viewed as an opportunity for the manifestation of new physics at short distance scales; i.e. in the ultraviolet (UV) regime. For example, let us suppose that there was some sort of interaction between different parts of the Dirichlet wall, but that such interactions could be ignored when the wall is the boundary of a large otherwise-smooth spacetime. This might be an intrinsically non-local interaction, or it might be the result of imposing particular boundary conditions at the wall for some local bulk field. Near $\tau=0$ the dynamics of such a system might then be well-modeled by our solutions above. But such interactions might lead to more dramatic effects near points at which our wall self-intersects. One might thus ask if such modifications to the above dynamical laws could prevent the bulk spacetime from shrinking to microscopic size. 

Since it is difficult to answer such questions in complete generality, let us restrict attention to new physics that respects causality, so that regions spacelike separated from our singularity cannot experience significant effects. The interesting question is then whether our spacetime contains any regions at all that lie in the causal future of our future singularity. If not, the causal new physics cannot prevent the spacetime from collapsing to microscopic scales.\footnote{Of course, even causal new physics could still influence the potential development (or avoidance) of a true mathematical singularity.}

In practice, this issue boils down to the question of whether our wall collision singularity is fully spacelike, or whether part of the singularity is timelike or null. 
We will calculate the result in section \ref{sec:ads3compactifications} for various cases with $d=3$.
However, before presenting explicit quantitative results, let us briefly reflect again on the motion of the boundary of our fundamental region relative to the static global coordinates $t,r,\phi$ {and make a heuristic argument about the causal character of these singularities in general dimension}. 

Since the velocity of this boundary vanishes at $t=0$, the motion is very slow at small $t$ (or equivalently at small $\tau$), though at each coordinate location $\chi$ it increases monotonically with $\tau$ up to $v(\chi) = \tanh \chi$ at $\tau=\pi/2$. 

\begin{figure}[h!]
    \centering
    \includegraphics[width=0.25\linewidth]{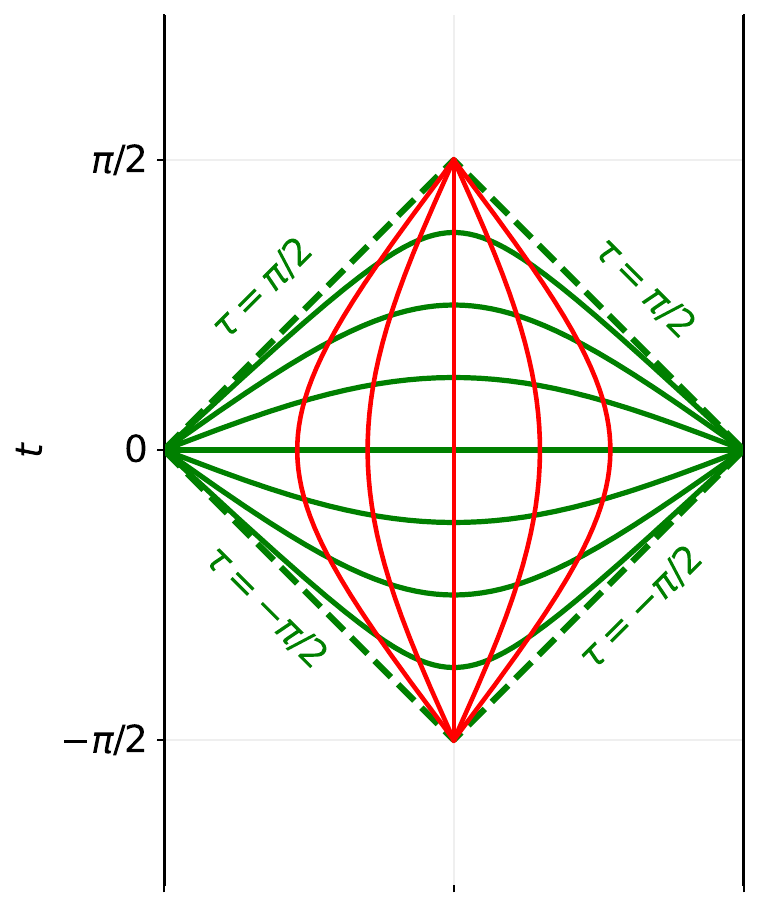}
    \caption{Constant $\tau$ surfaces (green) and constant $\chi$ surfaces (red) are shown on the conformal diagram of global AdS$_2$. Higher dimensional cases are similar and, in such cases, under time evolution points on the boundary of the fundamental domain also trace out curves of constant $\chi$. The central red line is both $\chi=0$ and $r=0$. Lines with constant $\chi=\chi_0$ and large $\chi_0$ reach small $r$ near $\tau=\pm \pi/2$, and they do so while moving at nearly the speed of light. }
    \label{fig:constanttausurfaces1}
\end{figure}

Furthermore, in the limit where the initial size of the fundamental domain is much larger than both the AdS scale $\ell$ and the radius $r_0$ of the Dirichlet wall\footnote{As we will see explicitly in section \ref{sec:ads3compactifications}, this limit requires taking the number of generators of $\Gamma$ to be large.}, the entire collision will take place near $\tau = \pi/2$. And since $\chi \rightarrow \infty$ in this limit at all points on the boundary of the fundamental domain, all parts of the collision process will take place at relativistic speeds $v\approx c$; see Figure \ref{fig:constanttausurfaces1}. A local model in the spirit of section \ref{subsec:flatcol} would thus involve the collision of two null planes. And since distinct null planes always intersect along a spacelike surface, in this limit the singularity should be everywhere spacelike. And since every constant $\chi$ line will hit the Dirichlet wall before $\tau = \pi/2$, this is again an end-of-time singularity where the volume of space shrinks to zero. This picture will be verified by direct calculation for the AdS$_3$ case in section \ref{sec:ads3compactifications}.

On the other hand, since two timelike planes whose orientations are related by any non-zero rotation intersect on a timelike line, singularities that arise from low-velocity
collisions can quickly become timelike. We will show by direct calculation in section \ref{sec:ads3compactifications} that this occurs when the Dirichlet wall nearly fills the fundamental domain at $\tau=0$.

\subsection{Dirichlet walls in AdS$_3$ cosmologies}
\label{sec:ads3compactifications}
We now display various details concerning spacetimes of the form described in section \ref{sec:genarg} above, specializing to the case of AdS$_3$ for simplicity. We will then consider a one-parameter generalization in section \ref{sec:conicalsingularities} characterized by the insertion of a conical singularity at $r=0$ (and thus which lies beyond our Dirichlet wall in the unphysical region of the spacetime). 

\subsubsection{ Constructing the solutions}
\label{subsec:constructads3}
We will first focus on the compactification of the two-dimensional hyperboloid into a genus $g$ Riemann surface. Since in comoving coordinates the boundaries of the fundamental domain do not change under time evolution, the desired quotient of AdS$_3$ is obtained by applying the same such compactification to all surfaces of constant $\tau$ \cite{Aminneborg:1997pz}. It will then be straightforward to insert our Dirichlet wall and to analyze the results.

The simplest example is to take the regular $4g$-gon as the fundamental domain and to identify opposite sides. The resulting spacetime is formed by taking the quotient with respect to the subgroup $\Gamma = \{\gamma_1, \gamma_2\dots \gamma_{2g}\},$ where $\gamma_1$ may be taken to be a boost in the $T_2,X_1$ plane in the embedding space $M^{2,d-1}$, and where $\gamma_k$ is related to $\gamma_1$ by a rotation through $\pi(k-1)/2g$ about the $X_1, X_2$ plane. In order to avoid conical singularities, the magnitude of the generators must be such that 
\begin{equation}
    \gamma_{2g} \dots \gamma_3^{-1} \gamma_2 \gamma_1^{-1} \gamma_{2g}^{-1} \dots \gamma_3 \gamma_2^{-1} \gamma_1 = 1.
\end{equation}
In particular, the above combination of successive transformations must act as the identity on any neighborhood of a vertex of the fundamental domain~\cite{Horowitz:1998xk, Balazs:1986uj}. An example with $g = 3$ is shown in Figure~\ref{fig:smallbig1}.

Alternatively, this construction can be thought of more geometrically: to avoid conical singularities, the interior angles of the fundamental domain must sum to $2\pi$. This condition is easily enforced on the fundamental domain by making use of the Gauss-Bonnet theorem. In particular, taking a quotient under any $\gamma_k$ identifies the planes
\begin{equation}
    T_2 = \pm \alpha \left(X_1 \cos{\theta} + X_2 \sin{\theta}\right),
\end{equation}
where $\theta = \pi(k-1)/2g$ and where $\alpha = \coth{\frac{\eta}{2}}.$ The intersections of such planes with the $\tau=0$ surface defines a $4g$-gon. From the Gauss-Bonnet theorem, the area must satisfy $A = 4 (g - 1)\pi$ in order to avoid a conical singularity. Computing the area by integrating the volume element of the induced metric and matching to the above result gives 
\begin{equation} \label{eq:alpha}
    \alpha = \coth{\frac{\eta}{2}} = \cos \left(\frac{\pi }{4 g}\right) \sqrt{\sec \left(\frac{\pi }{2 g}\right)}.
\end{equation}

Note that a non-singular surface requires $A>0$ and thus, since $g$ is an integer, $g\ge 2$. Furthermore, from \eqref{eq:alpha} we see that as $g$ grows, so does the boost parameter $\eta.$ But since $\alpha = \coth{\eta/2}$, our $\alpha$ decreases with $g$ and is bounded below by unity.
 Computing $\alpha$ for $g=2$ shows that for all allowed $g$ we have 
\begin{equation}
    1 < \alpha \leq \sqrt{\frac{1}{2} + \frac{1}{\sqrt{2}}}.
\end{equation}

Let us now introduce the desired Dirichlet wall. As mentioned above, the relevant surface is easily described in terms of the global coordinates \eqref{eq:global}. If we want a wall with circumference $L$ we simply set $r = L/2 \pi$. Indeed, in this coordinate system, the metric induced on the boundary is just:

\begin{equation}
    q_{ij} dx^i dx^j = - f\left(\frac{L}{2 \pi}\right) dt^2 + \left(\frac{L}{2 \pi}\right)^2 d \phi^2.
\end{equation}
In terms of the comoving coordinate system
\begin{equation} \label{eq:metric3d}
    ds^2 = - d\tau^2 + \cos^2 \tau \, (d\chi^2 + \sinh^2{\chi}\ d\phi^2),
\end{equation}
this boundary is given by
\begin{equation}
    L = 2 \pi \cos \tau \sinh \chi.
\end{equation} 

 For the spacetime to be non-singular at $\tau=0$, the desired wall must fit inside the $\tau=0$ fundamental region (without intersecting its boundary). This places an upper bound on $L$:
\begin{equation}\label{eq:Lbound}
    L < \frac{2 \pi}{\sqrt{\alpha^2-1}} = 2 \pi \sqrt{\cos \left(\frac{\pi }{2 g}\right)} \csc \left(\frac{\pi}{4 g}\right).
\end{equation}

\subsubsection{The causal character of the singularities}

The desired spacetimes are now given by the region that lies outside the Dirichlet wall (i.e., with $r \ge L/2\pi $) and inside our fundamental domain. Although the wall is chosen so that it does not intersect the boundary of the fundamental domain at $\tau=0$, as the universe contracts it will necessarily intersect this boundary at some $\tau < \pi/2$.

At such an intersection, points on the boundary cylinder become identified by the quotient. The intended Dirichlet boundary conditions thus become inconsistent at the identified points. Any attempt to enforce boundary conditions on this surface, with the identifications, leads to intersecting images of the Dirichlet boundary condition. The induced metric on the boundary cannot be reliably enforced, since at these intersections the boundary is no longer a manifold. In particular, the intersection of the desired constant-$r$ surface with a constant-$\tau$ slice is no longer a circle. Indeed, it is no longer given by any subset of the cylinder at all. Thus there can be no continuous extension of the solution beyond the above intersections that satisfies the desired boundary conditions. The intersections thus represent singularities at which the desired dynamics simply breaks down, and at which our spacetime must be cut off.

As discussed in section \ref{sec:genarg}, we would like to more fully characterize these spacetime singularities. In the future of the time-symmetric slice, the Dirichlet boundary first intersects the boundary of the fundamental domain at
\begin{equation} \label{eq:initialtime}
    \tau_i = \cos^{-1}\left( \frac{L}{2 \pi} \sqrt{\alpha^2 - 1}\right) = \cos ^{-1}\left(\frac{L}{2 \pi } \sin \left(\frac{\pi }{4 g}\right) \sqrt{\sec \left(\frac{\pi }{2 g}\right)}\right).
\end{equation}
This time is always less than $\pi/2$. It approaches $\pi/2$ as $L\rightarrow 0$, and approaches zero as $L$ reaches its maximum allowed value \eqref{eq:Lbound}. For fixed $L$, it increases with genus: increasing the genus pushed the formation of singularities to later times.

\begin{figure} [h!]
    \centering
    \includegraphics[width=0.9\linewidth]{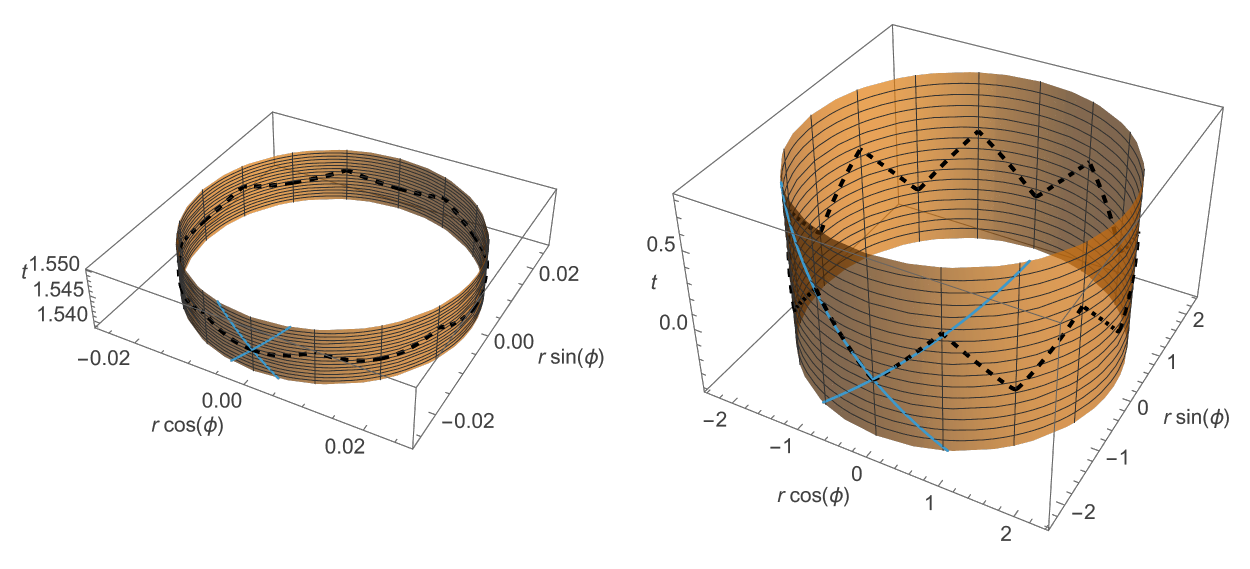}
    \caption{The future singularity on the Dirichlet wall is shown (dashed black line) for $g=2$ with $L$ satisfying \eqref{eq:spacelikeupperbound} (left), and violating \eqref{eq:spacelikeupperbound} (right). The particular values used are $L=0.157$ and $L=13.8$. Blue light cones are shown for reference, showing that the final moments of the singularity are respectively spacelike and timelike as expected (though the final `velocity' of the singularity, defined with respect to lines of constant $r, \phi$ on the right is still over 0.9$c$). In all cases the singularity is a smooth spacelike curve when it first forms.}
    \label{fig:smallandlargewallsingularity}
\end{figure}

At sufficiently late times, the entire boundary becomes singular. This occurs when the boundary cylinder reaches the vertices of the fundamental domain. At that moment, the spacetime ceases to exist: since the fundamental domain lies entirely within the boundary cylinder, there is no {remaining} spacetime outside of the cylinder. This final time is
\begin{equation} \label{eq:finalsingularity}
    \tau_f = \cos^{-1}\left(\frac{L}{2 \pi} \sqrt{\alpha ^2 \cos ^2\left(\frac{\pi}{4 g}\right)-1}\right) = \cos ^{-1}\left(\frac{L}{2 \pi } \sin ^2\left(\frac{\pi }{4 g}\right) \sqrt{\sec \left(\frac{\pi }{2 g}\right)}\right).
\end{equation}
Between these two times, the singularities lie on the curves, in $(\tau, \chi, \phi)$ coordinates \eqref{eq:metric3d},
\begin{equation} \label{eq:singularities}
    \left(\tau,~\sinh ^{-1}\left(\frac{L}{2 \pi} \sec \tau \right),~\frac{n \pi}{2 g}\pm \cos^{-1}\left(\frac{2 \pi}{\alpha L}\sqrt{\cos ^2\tau+\left(\frac{L}{2 \pi}\right)^2}\right) \right),
\end{equation}
where $n = 0, 1 \dots 4g - 1$ (see Fig.~\ref{fig:smallandlargewallsingularity}). There are additional singularities given by time reflection.

The constructed spacetime is akin to a cosmology with initial and final singularities, and where the singularities cut off the boundary wall as well as cutting off the bulk.

As discussed in section \ref{sec:genarg}, it is of interest to determine whether our singularity is fully spacelike, or whether part of the singularity is timelike or null. There it was argued in particular that the singularity is always spacelike when it first forms, and that it should remain fully spacelike in spacetimes where the fundamental domain at $\tau=0$ is much larger than both the AdS scale ($\ell=1$) and the Dirichlet wall. Note in particular that the former requirement implies $g\gg 1$. On the other hand, for walls that nearly fill the $\tau=0$ fundamental domain it was plausible that the initially-spacelike singularity quickly becomes timelike. We will see below that this is indeed the case.

To proceed, we need only compute the norm of the tangent to the curve \eqref{eq:finalsingularity}. We find the singularity to be everywhere spacelike when $L$ satisfies
\begin{equation} \label{eq:spacelikeupperbound}
    L < 2 \pi \sqrt{\frac{1-\alpha ^2 \sin ^2\left(\frac{\pi }{4 g}\right)}{\alpha ^2-1}} = 2 \pi \sqrt{\cos ^2\left(\frac{\pi }{4 g}\right) \cot ^2\left(\frac{\pi }{4 g}\right)-1}
\end{equation}
This is in fact a stronger result than that expected from the general arguments of section \ref{sec:genarg}, as our singularities are everywhere spacelike for small $L$ at {\it any} fixed $g\ge 2$ rather than just in the limit of large $g$. In contrast, when $L$ is too large to satisfy \eqref{eq:spacelikeupperbound} but still below the maximum \eqref{eq:Lbound}, the singularity is spacelike when it first forms but then becomes timelike at later times.

Since the Gauss-Bonnet theorem ensures that the initial area $A$ of the fundamental domain diverges as $g\rightarrow \infty$, for any $L$ there is a spacetime where the wall self-intersection curve is spacelike. In particular, in this limit we have $\alpha \rightarrow 1$ so that the right-hand-side of \eqref{eq:spacelikeupperbound} diverges. Thus for any flat metric on the cylinder there is a solution where adding causal new UV physics cannot avoid collapse of the entire space to microscopic size {as long as the new UV physics is active only in regions where curvatures are high or where the distance between two walls has become small}.

\subsubsection{Brown-York stress energy tensor}
\label{sec:brownyork}
Since our Dirichlet wall is static, there is a conserved energy for our spacetimes. Indeed, their energies can be meaningfully compared with boxed BTZ black hole spacetimes with corresponding Dirichlet walls. We will see that (for fixed genus) decreasing the energy increases the time to the singularity and makes the singularity more spacelike.
 
On general grounds the energy can be computed using the 
 Brown-York stress-energy tensor defined by
\begin{equation}
    T_{ij} = \frac{-2}{\sqrt{-q}} \frac{\delta S}{\delta q^{ij}},
\end{equation}
where $q_{ij}$ is the induced metric on the boundary and $S$ is the gravitational action. We use Latin indices to indicate coordinates on the boundary. For a Dirichlet boundary this is
\begin{equation} \label{eq:brown-york-definition}
    T_{ij} = -\frac{1}{8 \pi }\left(K_{ij} - K q_{ij}\right),
\end{equation}
where $K_{ij}$ is the extrinsic curvature of the boundary (with $K$ its trace) and we have set is Newton's constant $G=1$. 

The conserved energy is then
\begin{equation}
    E = \int_C T_{ij} \xi^i n^j dC,
\end{equation}
where $C$ is a spacelike cross-section of the Dirichlet boundary, {$dC$ is the induced volume element (here a length) on $C$}, $\xi^i$ is the timelike Killing vector, and $n^j$ is the unit normal vector to $C$ within the Dirichlet wall. If we take $C$ to be a surface of constant global time $t$, then $\xi^i = n^i$ and only the time-time component of $T_{ij}$ enters this expression. This component is independent of the time-time component of $K_{ij}$ and depends only on the trace of its spatial components. This is consistent with the familiar feature that the ADM energy at infinity is determined by the induced metric on a Cauchy slice.

For a general static metric of the form
\be 
ds^2 = -f(r) dt^2 + \frac{dr^2}{f(r)} + r^2 d\phi^2
\ee
the extrinsic curvature on a constant $r=r_0$ surface (using the outward pointing normal) is $g^{\phi\phi} K_{\phi\phi} = f^{1/2}(r_0)/r_0$. Assuming $\phi$ has period $2\pi$, this yields an energy $E = -f^{1/2}(r_0)/4$, so for the BTZ black hole (with $f(r) = r^2 -M$)
\be
E_{BTZ} = -\frac{(r_0^2-M)^{1/2}}{4} \qquad {\rm for \ outward \ normals.}
\ee
So for all walls outside the horizon, the energy is negative\footnote{In the limit $r_0 \rightarrow \infty$ this energy diverges and one usually regulates it in a way that leaves a positive finite energy. Since we get a finite energy for finite $r_0$, there is no need for any ``counterterm subtraction".}.

For the spacetimes constructed in section \ref{subsec:constructads3}, we are interested in the region outside our wall, so we need to compute the extrinsic curvature with respect to the {\it inward} normal. This changes its sign relative to the results above. Writing the energy in terms of the length $L$ of the boundary then yields
\be
E_{BTZ} = \frac{1}{4}\left[\frac{L^2}{4\pi^2} -M\right]^{1/2} \qquad {\rm for \ inward \ normals.}
\ee
In fact, the solutions of section \ref{subsec:constructads3} have pure $\AdS_3$ near our boundary wall which corresponds to $M = -1$. Thus we find
\begin{equation}
    E = \frac{1}{8\pi}\left[{L^2}+{4\pi^2}\right]^{1/2}
\label{eq:E1}\end{equation}

Note that \eqref{eq:E1} is always positive,
and so distinguishes our solutions with end-of-time singularities from boxed BTZ spacetimes. However, it is also independent of the genus $g$ (so long as $g$ is large enough to allow a wall of length $L$). As a result, it does not - by itself - distinguish solutions with everywhere-spacelike singularities (which occur for large $g$ with fixed $L$) from those where the singularities eventually become timelike (which occur when $L$ is near the maximum allowed value for a given $g$). 

\subsection{A one-parameter family of vacuum solutions}
\label{sec:conicalsingularities}

The above construction of vacuum solutions with Dirichlet walls admits a simple one-parameter generalization. To explain this generalization, let us first note that the AdS cosmologies (without Dirichlet walls) can be modified by including a conical defect at the origin $r=0$ of our global AdS coordinates. We may again take the global AdS metric to be given by \eqref{eq:metric3d}, but now with $0\le \phi < 2\pi-\delta$ for some $\delta < 2\pi$. Note in particular that we allow conical excesses with $\delta < 0$.

If this is the only conical singularity, using the Gauss-Bonnet theorem at $t=0$ (equivalently, at $\tau=0$) now gives 
\begin{equation}
\label{eq:GBwdelta}
    A = \pi (4 g - 2) - (2 \pi - \delta) = 4 \pi(g - 1) + \delta.
\end{equation} The resulting fundamental domain no longer tiles the hyperbolic plane, though it could now be considered to tile a plane which is locally hyperbolic at generic points but which contains an infinite set of regularly spaced conical singularities of deficit $\delta$. As a result of \eqref{eq:GBwdelta}, the parameter $\alpha$ associated with identifying opposite boundaries of the associated domain now depends on both $g$ and $\delta$:
\begin{equation} \label{eq:alphadeficitorigin}
    \alpha = \cos \left(\frac{\pi }{4 g}\right) \sqrt{\sec \left(\frac{\delta }{8 g}\right) \sec \left(\frac{4 \pi - \delta}{8 g}\right)}.
\end{equation}
As before, the coordinate locations of the above features are $\tau$-independent when expressed in comoving coordinates so that physical effects enter only through the scale factor $\cos \tau$ in \eqref{eq:metric}.

It is now straightforward to again introduce a Dirichlet wall whose induced metric (before the singularity forms) is that of a static cylinder of circumference $L$ by locating the wall on a surface $r=constant$. Again, we keep only the part of the spacetime outside the wall. In particular, inserting the wall will excise the conical singularity from the physical spacetime. Nevertheless, due to the restricted range of $\phi$ we now find $r = \frac{L}{2\pi - \delta}$, which in comoving coordinates yields
\begin{equation}
    \frac{L}{2 \pi - \delta} = \cos \tau \sinh \chi.
\end{equation}

The singularities again develop along the boundary when this surface collides with the boundaries of the fundamental domain. In particular, the singularity now first develops at
\begin{equation} \label{eq:initialtimedeficit}
    \tau_i = \cos ^{-1}\left(\frac{L \sqrt{\alpha ^2-1} }{2 \pi - \delta}\right) = \cos ^{-1}\left(\frac{L}{2 \pi -\delta } \sqrt{\cos ^2\left(\frac{\pi }{4 g}\right) \sec \left(\frac{\delta }{8 g}\right) \sec \left(\frac{4 \pi - \delta}{8 g}\right)-1}\right),
\end{equation}
and the time at which the entire boundary has become singular is
\begin{equation} \label{eq:finaltimedeficit}
\begin{split}
    \tau_f &= \cos ^{-1}\left(\frac{L}{{2 \pi -\delta }} \sqrt{\alpha ^2 \cos ^2\left(\frac{2 \pi - \delta}{8 g}\right)-1}\right)\\
    &= \cos ^{-1}\left(\frac{L}{2 \pi -\delta } \sqrt{\sin ^2\left(\frac{\pi }{4 g}\right) \sin ^2\left(\frac{2 \pi - \delta}{8 g}\right) \sec \left(\frac{\delta }{8 g}\right) \sec \left(\frac{4 \pi - \delta}{8 g}\right)}\right).
\end{split}
\end{equation}
These results of course 
reduce to~\eqref{eq:initialtime} and ~\eqref{eq:finalsingularity}
for $\delta = 0$. At intermediate values of $\tau$ the singular points are at 
\begin{equation} \label{eq:singularitydeficit}
    \left(\tau ,\sinh ^{-1}\left(\frac{L \sec (\tau )}{2 \pi -\delta }\right), \frac{n \pi}{2 g} \pm \cos ^{-1}\left(\frac{(2 \pi -\delta )}{\alpha L} \sqrt{\frac{L^2}{(2 \pi - \delta)^2}+\cos ^2(\tau )}\right)\right),
\end{equation}
where again $n = 0, 1 \dots 4g - 1$. 

\begin{figure}
    \centering
    \includegraphics[width=0.5\linewidth]{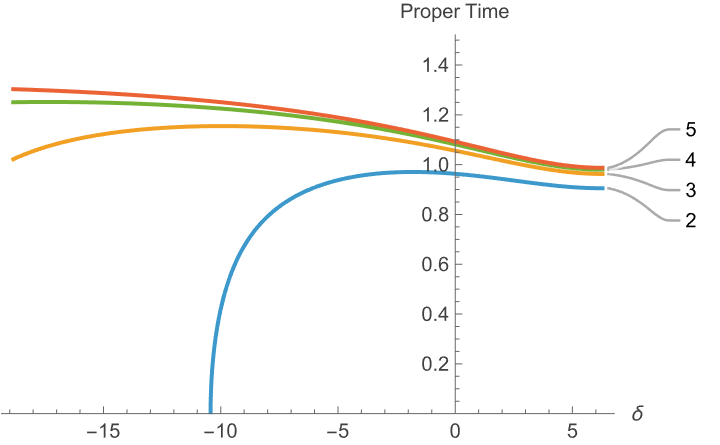}
    \caption{The proper time along the wall from the time symmetric surface to the point at which the singularity first forms as a function of the conical deficit $\delta$, shown for $g = 2, 3, 4, 5,$ with $L = 2 \pi$. For increasing (positive) $\delta$ this time decreases since the radial coordinate of the wall increases to keep the circumference constant. For $\delta < 0$, the proper time to the singularity vanishes when $L = 2\pi$ saturates the geometrically derived maximum on the circumference of the wall given by Eq.~\eqref{eq:Lbound2}.}
    \label{fig:formationtime}
\end{figure}

In Figure~\ref{fig:formationtime}, we plot the proper time along the wall from the time symmetric surface to the point at which the singularity first forms as a function of the conical deficit $\delta$. The time decreases for increasing (positive) $\delta$: this can be understood by considering the radial coordinate of the wall, which increases to keep the circumference constant and thus decreases the time to the singularity. However, as $\delta$ becomes more negative, the radial coordinate of the innermost point of the fundamental domain decreases, which also decreases the proper time to the singularity. For sufficiently negative conical deficit, this effect outweighs the former. The proper time to the singularity vanishes when $L$ saturates the geometrically derived maximum on the circumference of the wall given by Equation~\eqref{eq:Lbound2}.

\begin{figure}
    \centering
    \includegraphics[width=0.5\linewidth]{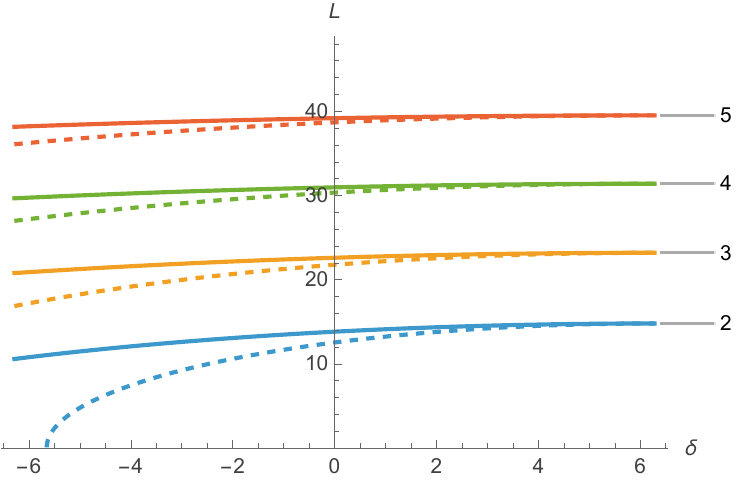}
    \caption{Regions of parameter space featuring fully spacelike vs. partially timelike singularities.
    Parameters below the dashed lines for a given genus give singularities that are everywhere spacelike. Solid lines indicate values of $L$ for which the wall just touches the boundaries of the fundamental domain at $\tau=0$, so non-singular initial data exists below this line. For fixed $L,\delta$, there is always a sufficiently high genus such that the singularities remain everywhere spacelike.}
    \label{fig:tlsingularities}
\end{figure}

As before, requiring that there be no singularities in the time symmetric surface implies only an upper bound on the size of the Dirichlet wall given by 
\begin{equation}\label{eq:Lbound2}
    L < \frac{2 \pi - \delta}{\sqrt{\alpha^2-1}} = (2 \pi -\delta) \sqrt{\cos ^2\left(\frac{\pi }{4 g}\right) \csc ^2\left(\frac{2 \pi - \delta}{8 g}\right)-1}.
\end{equation}
But we again find the singularity to be everywhere spacelike for small $L$, and in particular for 
\begin{equation} \label{eq:spacelikeupperbound2}
    L < \frac{(2 \pi -\delta )}{\sqrt{\alpha ^2-1}} \sqrt{1-\alpha ^2 \sin ^2\left(\frac{2 \pi - \delta}{8 g}\right)} = (2 \pi -\delta ) \sqrt{\cos ^2\left(\frac{\pi }{4 g}\right) \left(\csc ^2\left(\frac{2 \pi -\delta }{8 g}\right)-1\right)-1}.
\end{equation}
These bounds are shown in Figure~\ref{fig:tlsingularities}. As the genus increases, so does this bound. Thus for fixed circumference $L$ and deficit angle $\delta$, there always exists a large enough genus such that the singularities remain everywhere spacelike. The bound also increases with conical deficit: a larger deficit angle allows the introduction of a larger Dirichlet boundary wall while keeping the singularity everywhere spacelike. 

Finally, we follow section~\ref{sec:brownyork} in computing the extrinsic curvature, Brown-York stress energy tensor, and conserved energy to find 
\begin{equation}
    E = \frac{1}{8 \pi }\sqrt{L^2 + \left(2 \pi - \delta \right)^2}.
\end{equation}
The conserved energy is again strictly positive, for all allowed values of the parameters, and is again independent of $g$.
In Figure 7 we compare the energy to when the singularity is everywhere spacelike. It is clear that lower energy solutions {tend to have } spacelike singularities.

A further generalization is studied in Appendix~\ref{sec:pointmass}. There we consider the insertion of masses into the above solution at the point defined by the `corners' of the fundamental domain. The expressions for the various quantities are generally the same as those in this section when written as functions of $\alpha$. In particular, the conserved energy is independent of the point mass and only depends on the conical singularity at the origin. This independence is expected, since adding the point mass does not change the solution near the Dirichlet wall and thus cannot affect the Brown-York stress tensor.

\begin{figure}
    \centering
    \includegraphics[width=0.5\linewidth]{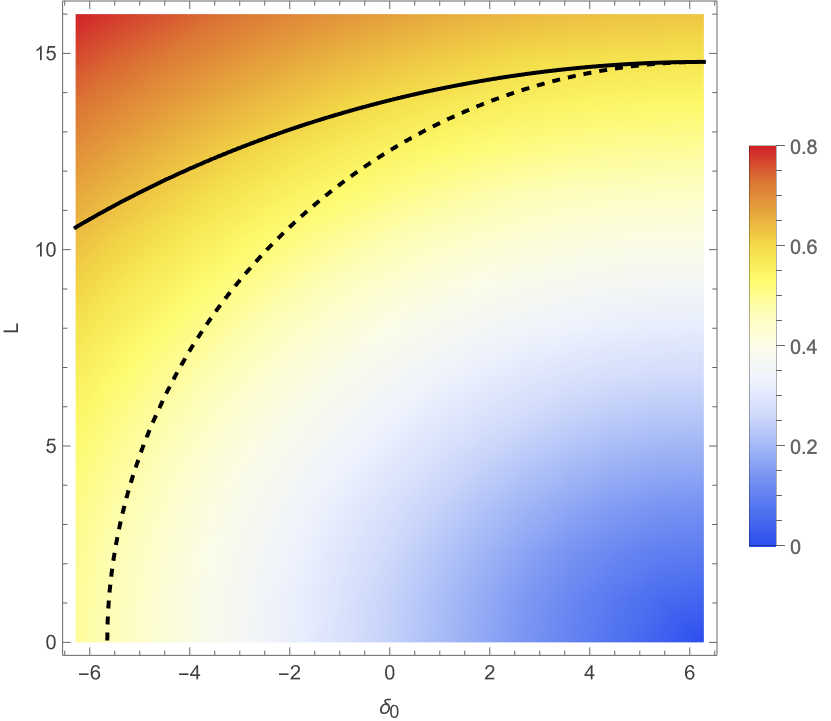}
    \caption{ The conserved energy $E$ as a function of conical deficit $\delta$ and boundary size $L$ for the case $g=2$. Red/blue shading indicates {large/small $E$}. The solid black line is the upper bound on $L$ from the absence of singularities at $\tau=0$. The dashed line marks the threshold below which the singularities are everywhere spacelike. Notice that lower energy solutions tend to have spacelike singularities.}
    \label{fig:energy}
\end{figure}

\subsection{Changing the sign of $E$}
\label{sec:curvature}

As noted above, our end-of-time solutions can be distinguished from more familiar boxed-black-hole solutions by the sign of the conserved energy $E$. This naturally leads to the question of whether one might be able to regard the $E>0$ and $E<0$ cases as separate theories so that, in particular, the $E<0$ theory might be free of our end-of-time solutions.\footnote{Here we use the precise definition of energy introduced above which defines a particular notion of zero-energy solutions. However, we are also free to add finite counterterms to the definition of the stress-energy tensor in Equation~\eqref{eq:brown-york-definition}, which can shift the total energy by a constant $E_0$. In that case the two cases are simply $E>E_0$ and $E<E_0$.}

However, at least in the presence of appropriate matter fields, the $E>0$ and $E<0$ cases are continuously connected through $E=0$. As an example, let us consider the Einstein static universe $S^{d-1}\times {\mathbb R}$ supported by the usual combination of positive cosmological constant $\Lambda$ and pressureless dust.
This solution has a $Z_2$ symmetry that leaves invariant the equatorial $S^{d-2}$ on the $S^{d-1}$. Placing our wall on that equatorial sphere thus gives a static Dirichlet wall with the desired induced metric and with vanishing extrinsic curvature (so that $E$ then vanishes as well). This solution also clearly interpolates between $E>0$ and $E<0$ solutions in which the wall sits on one side of the equator or the other.

While the above solution requires a precise relation between the choice of $\Lambda$ and the radius of the wall, this turns out to be a consequence of seeking a simple solution with exact time-translation symmetry. For example, for some given value of $\Lambda$, let us now take the energy density $\rho$ in the pressureless dust to be slightly less than the value associated with the Einstein Static Universe. Then there is a solution (for now, with no wall) which has a time-symmetric slice (which we will call $\tau=0$) on which the volume of the $S^{d-1}$ is minimal and away from which this volume grows in both directions. Since the volume grows only very slowly, there are timelike surfaces of induced metric $S^{d-2} \times {\mathbb R}$ that start on the equatorial $S^{d-1}$ at $\tau=0$ but have some non-zero initial velocity, and which thus drift to one side as time increases. Away from $\tau=0$ we will need to correct the spacetime for the fact that the wall now collides with various dust particles and changes their motion, but such corrections will be small near $\tau=0$ so corrected solutions will indeed exist at least for sufficiently small times. In these solutions the size of the wall's $S^{d-2}$ is no longer precisely determined by the cosmological constant $\Lambda$.

Of course, if the boundary conditions on the wall are fully invariant under time-translations, there can be no solution along which the sign of $E$ actually changes with time. However, such solutions are easily found if we drive the system by imposing time-dependent boundary conditions {(unless this leads to a different kind of singularity through a failure of the system to be well-posed)}. In the above context of $\Lambda >0$ solutions with dust, it turns out to be sufficient to simply impose `transparent' boundary conditions that allow the dust particles to fly free through the wall (which would be described as absorption of matter by the wall in a more realistic theory, and which would then require time-dependent boundary conditions for the matter fields). In particular, if we follow the construction of the above paragraph with such transparent boundary conditions there above-mentioned corrections vanish and we can easily compute $E(t)$. 

Note that the positive cosmological constant and pressureless dust used in the above construction need not be fundamental constituents of the relevant theory. Indeed, it is enough to consider any theory with a scalar field $\phi$ whose potential $V(\phi)$ becomes positive at some $\phi$. Considering initial data where gradients of the field happen to be small would then reproduce the effect of a positive cosmological constant near the initial surface $\tau=0$. And adding excitations of the scalar field can then approximate the effects of radiation, which can be used in place of the above pressureless dust to make an Einstein Static Universe or approximations thereto. 

If one wishes to consider a completely different construction which again allows us to drive the system from $E<0$ to $E>0$, one might consider any spherical cosmology (topologically $S^{d-1} \times {\mathbb R}$) with $\Lambda \le 0$ and with additional matter that allows the spacetime to have a moment of time-symmetry. Associated with the fact that
simple such spacetimes collapse, they will not have {\it timelike} surfaces with a static $S^{d-2} \times {\mathbb R}$ metric that crosses the equator of the $S^{d-1}$. However, one can readily find such surfaces with time-dependent induced metrics, and in particular which become static for some time periods both before and after crossing the equator. As a result, one can again construct {\it driven} solutions along which $E(t)$ changes sign. In particular, such solutions can approximate a familiar boxed black hole spacetime at early times while featuring our end-of-time singularities at late times.

{Before ending this section, we should warn the reader that the assumptions under which An and Anderson established that the Dirichlet initial boundary value problem are well-posed fail to hold in the solutions constructed above. The first issue is that the results of \cite{anWellposedGeometricBoundary2025a} were proven only for the vacuum Einstein equations (with vanishing cosmological constant). However, even if they could be generalized to the theories discussed above,} the theorem of \cite{anWellposedGeometricBoundary2025a} for Dirichlet boundary conditions was proven under the assumption that the signature of the Brown-York stress tensor $T_{ij}$ on the Dirichlet wall agrees with the signature of the induced metric $\gamma_{ij}$. The proof also assumed the vector field $\partial_t$ along which one wishes to time-evolve to be future-directed and timelike as defined by both $\gamma_{ij}$ and $T_{ij}$ (with the latter now interpreted as an additional metric). As described in footnote \ref{foot:AA}, this can be generalized to the case where $-T_{ij}$ satisfies the above conditions (instead of $T_{ij}$) using precisely the same proof. But the argument fails at the moment where $T_{tt}$ vanishes. 

This failure may itself have interesting implications. However, from the above examples we see that it will necessarily occur in physically natural settings. Indeed, another natural setting in which this assumption fails, and in fact in which the sign of $T_{tt}$ at points on the boundary would be expected to oscillate locally in time, is easily constructed by first considering linearized gravitational waves in Minkowski space. Using the method of images, we can of course construct waves that satisfy Dirichlet boundary conditions on any timelike plane, and in particular we can use a superposition of two plane waves. In empty Minkowski space, the Brown-York stress tensor of such a plane vanishes due to the $Z_2$ reflection symmetry that preserves the plane. So in the presence of our plane waves (say, with frequency $\omega$) the Brown-York stress tensor clearly oscillates in time with the same frequency $\omega$. At a given location on the wall, a given component thus spends half of each cycle with one sign and half of each cycle with the opposite sign\footnote{If the plane waves propagate along the direction normal to the wall, and if we take the wall to be at rest in a given inertial frame, then that component $T_{tt}$ of the Brown-York stress tensor vanishes identically due to the fact that gravitational waves are transverse-traceless. However, $T_{tt}$ is non-zero for properly-polarized waves that are incident at an oblique angle or, equivalently, if we transform the normal-incidence case to a new inertial frame via a boost along e.g. either the $x$ or $y$ directions in the case of ``$+$" polarization.}.

Of course, since a flat wall in Minkowski space has $T_{ij}=0$, it already fails the An-Anderson condition. But this degeneracy can be removed by adding a slight curvature to the wall, for example by taking it to be $S^{d-2} \times {\mathbb R}$ in $d$-dimensional Minkowski space and taking the $S^{d-2}$ to have a large radius $R$. Then for $d \ge 4$, if we choose the overall sign of $T_{ij}$ to be that associated with taking our spacetime to be the region {\it inside} the wall, we obtain a Brown-York stress tensor $T^{Mink}_{ij}$ that satisfies the An-Anderson condition. 

Let us now add gravitational waves as above, say with an amplitude $A$ small enough to use the linearized approximation but still with $A{\omega} \gg 1/R$. If we take the gravitational waves to be localized in the interior (so that they are very small in the neighborhood of the wall) at $t=0$, then again An-Anderson is satisfied at that time. But when suitably general such waves hit the wall, every component of $T_{ij}$ will begin to oscillate with amplitude larger than any component of $T^{Mink}_{ij}$. Thus the An-Anderson condition is necessarily violated when the radiation arrives and, in particular, the local energy density $T_{tt}$ at points on the wall should oscillate in sign. 

It thus seems physically important to ask how solutions behave when the An-Anderson condition is violated. {The above examples illustrate that there are two logical possibilities. The first is that the theory remains well-posed throughout the deformations discussed above so that the $E>0$ and $E<0$ solutions live in the same connected component of a single well-posed theory. In that case, the usual boxed black hole solutions can be deformed to solutions that feature our end-of-time singularities. The second option is simply that the theory ceases to be well-posed at some finite point along this deformation. But at such a point it would be natural to say that the theory again features a kind of singularity and, in particular, that the dynamics of the theory have become singular. Thus in either case singularities arise under finite deformations of boxed black hole initial data. }

\section{Flat walls in the BTZ black hole}
\label{sec:WallinBTZ}

In the previous section we argued that gravity with Dirichlet walls can lead to end-of-time singularities. In addition, we argued that such singularities can result even from solutions that, at some initial time, are very similar to standard boxed black hole solutions formed by taking a Dirichlet wall to surround a Schwarzschild black hole. However, the construction involved either driving the system with time-dependent boundary conditions or using either a positive cosmological constant or a scalar field proxy. We therefore now construct boxed black hole solutions with time-independent boundary conditions that lead to similar singularities by breaking spherical symmetry and studying a region of the wall on which the local Brown-York energy density $T_{tt}$ is positive even when the total energy has $E<0$.

One way to construct such solutions with Dirichlet walls is to start with a known wall-free solution and to look for a surface with the prescribed induced metric. This is particularly useful in 2+1-dimensional vacuum solutions as then there are no local degrees of freedom with which the wall can interact, so the addition of the wall can induce no back-reaction even in the absence of spherical symmetry.

If the known solution is a black hole and the surface enters the horizon, then evolution must stop when the surface hits the singularity. Note that there is no problem with a timelike surface (our Dirichlet wall) crossing a null surface (a black hole horizon). 

We therefore now study two-dimensional surfaces with induced static $S^1 \times {\mathbb R}$ metrics in a non-rotating BTZ black hole. It is easy to construct flat surfaces with topology $R^2$: If we write the BTZ black hole with $\ell=r_+=1$ in ingoing null coordinates
\begin{equation}
ds^2 = -(r^2 -1) dv^2 + 2dv dr + r^2 d\phi^2
\ee
then the surfaces $v = \pm \phi$ have induced metric 
\be 
ds^2 = dv^2 + 2dv dr
\ee 
and hence are flat timelike surfaces. Since all values of $r$ are included, these surfaces clearly enter the black hole. 

It is also easy to find surfaces with induced static metrics $S^1 \times {\mathbb R}$ that enter the black hole when the size of the $S^1$ is sufficiently small. To do so, we need only choose a timelike geodesic that falls into the black hole and to note that, due to the lack of local degrees of freedom in 2+1 dimensions, in small enough neighborhoods of that geodesic the metric will be isometric to that of a neighborhood of the central ($r=0$) geodesic in global AdS$_3$. As a result, that region contains a cylindrical timelike surface with static induced $S^1 \times {\mathbb R}$ metric for at least some non-zero interval of proper time. However, since the surface falls into the black hole, the surface will again encounter the future singularity in finite time.

However, the above cases are both topologically distinct from the familiar boxed black hole solutions. To have a solution that can be viewed as a {deformation of a} boxed black hole, we need the Dirichlet wall to have topology $S^1 \times R$ and for it to enclose the black hole. If we start with a constant radius circle on a static slice, {and if we give it vanishing initial velocity,} the corresponding flat surface stays at constant radius and never crosses the horizon. However, we show below that more general surfaces can enter the horizon.\footnote{{We thank ChatGPT for help in obtaining the following result.}}

Let us consider the non-rotating BTZ geometry in the Kruskal-like coordinates
\begin{equation}
 ds^2=\frac{-dT^2+dX^2+\left(1-\frac{T^2-X^2}{4}\right)^2d\phi^2}
 {\left(1+\frac{T^2-X^2}{4}\right)^2}.
 \label{eq:BTZ95TXphi}
\end{equation}
We embed a timelike surface with coordinates $(\tau,\sigma)$ via
$X^\mu(\tau,\sigma)=(T(\tau,\sigma),X(\tau,\sigma),\phi(\tau,\sigma))$.

We begin with a smooth closed curve $C$ in the static slice $T=0$,
\begin{equation}
 T=0,\qquad X=X_0(\sigma),\qquad \phi=\sigma,
 \label{eq:initial-curve-GC}
\end{equation}
where $X_0$ is $2\pi$-periodic. 
Define
\begin{equation}
 c_0=1-\frac{X_0^2}{4},\qquad
 q_0=1+\frac{X_0^2}{4},\qquad
 d_0=X_0'^2+q_0^2 ,
 \label{eq:cqD-GC}
\end{equation}
and assume $c_0>0$. Proper length $s$ along $C$ is then given by
\begin{equation}
 ds=\frac{\sqrt{d_0}}{c_0}\,d\sigma .
 \label{eq:proper-length-GC}
\end{equation}
On the $T=0$ slice let $e$ be the unit tangent to $C$, and choose the unit (outward)
normal $m$ within this slice to be
\begin{align}
 e&=\frac{c_0}{\sqrt{d_0}}
 \left(X_0'\,\partial_X+\,\partial_\phi\right),
 \nonumber\\
 m&=\frac{c_0}{\sqrt{d_0}}
 \left(q_0\,\partial_X-\frac{X_0'}{q_0}\,\partial_\phi\right).
 \label{eq:em-GC}
\end{align}
By time reflection symmetry, the proper acceleration of $e$ must be proportional to $m$. We therefore define the signed
acceleration $a(s)$ of the curve by
\begin{equation}
 e^\nu\nabla_\nu e^\mu=a(s)m^\mu,
 \label{eq:curve-accel-def}
\end{equation}
where $\nabla_\nu$ is the BTZ covariant derivative.

Our goal is to construct a flat timelike cylinder passing through $C$. One might think that the simplest approach would be to look for a time-symmetric surface. However, this is not always possible. {Indeed, we } now show that there is an obstruction to constructing a time-symmetric flat surface if $C$ includes a maximum far from the black hole and a minimum close to the horizon.

Let $t=c_0\partial_T$ denote the future-directed unit timelike normal to a static slice of the BTZ geometry. If the wall is required to be strictly time symmetric at $C$, its future-directed unit tangent normal to $C$ within the wall is $u^\mu=t^\mu$, while the unit {outward} normal to the wall is $n^\mu=m^\mu$. Let $K_{\mu\nu}$ denote the extrinsic curvature (defined with respect to the outward normal). We write its components in the orthonormal frame $(u^\mu,e^\mu)$ as
\begin{equation}
 K_{00}=K_{\mu\nu}u^\mu u^\nu,\qquad
 K_{01}=K_{\mu\nu}u^\mu e^\nu,\qquad
 K_{11}=K_{\mu\nu}e^\mu e^\nu .
 \label{eq:K-components-GC}
\end{equation}
Time symmetry
implies
\begin{equation}
 K_{01}=0,\qquad K_{11}= -a(s).
 \label{eq:K-timesym-GC}
\end{equation}
Since BTZ is locally $\mathrm{AdS}_3$ with unit curvature radius, the Gauss
equation gives the scalar curvature ${\cal R}[q]$ of the induced metric $q$ as
\begin{equation}
 {\cal R}[q]=-2+2\left(K_{01}^2-K_{00}K_{11}\right).
 \label{eq:Gauss-scalar-GC}
\end{equation}
A flat wall therefore requires
\begin{equation}
 K_{01}^2-K_{00}K_{11}=1.
 \label{eq:Gauss-flat-GC}
\end{equation}
For time-symmetric data this reduces to
\begin{equation}
 K_{00}=\frac{1}{a(s)}.
 \label{eq:timesym-K00-GC}
\end{equation}
Thus a smooth strictly time-symmetric flat wall cannot pass through an initial curve for which $a(s)$ vanishes. Indeed, returning to \eqref{eq:Gauss-scalar-GC}, we see that, at such a point, both $K_{01}$ and $K_{11}$ vanish so the induced scalar curvature is forced to be ${\cal R}[q]=-2$ for any $K_{00}$.

For any closed curve in the $T=0$ surface that has a maximum far from the horizon and a minimum near the horizon, the acceleration must change sign somewhere. This is clear since a curve at constant $X$ has acceleration in the $-\partial_X$ direction. On the other hand, the $X=0$ horizon is a geodesic, so a curve that has a minimum near the horizon has acceleration in the $+\partial_X$ direction.

We can avoid this obstruction by allowing the wall to have a small region of nonzero initial velocity. Let us introduce a rapidity $\chi(s)$ and take the timelike
tangent and spacelike normal to the wall along $C$ to be
\begin{equation}
 u^\mu=\cosh\chi\,t^\mu+\sinh\chi\,m^\mu,
 \qquad
 n^\mu=\sinh\chi\,t^\mu+\cosh\chi\,m^\mu.
 \label{eq:boosted-frame-GC}
\end{equation}
The corresponding first time derivatives of the embedding, when $\tau$ is
proper time along $u^\mu$, are
\begin{align}
 \left.\partial_\tau T\right|_{\tau=0}
 &=c_0\cosh\chi,
 \nonumber\\
 \left.\partial_\tau X\right|_{\tau=0}
 &=\frac{c_0q_0}{\sqrt{d_0}}\sinh\chi,
 \nonumber\\
 \left.\partial_\tau\phi\right|_{\tau=0}
 &=-\frac{c_0X_0'}{q_0\sqrt{d_0}}\sinh\chi.
 \label{eq:initial-velocity-GC}
\end{align}
A short calculation using the fact that the $T=0$ slice is totally geodesic
gives 
\begin{equation}
 K_{11}=-a\cosh\chi,
 \qquad {\rm and}
\qquad K_{01}=- \frac{d\chi}{ds}.
 \label{eq:K11K01-GC}
\end{equation}
It also gives the acceleration of $C$ as a curve in the induced two-dimensional metric. If we denote the intrinsic covariant derivative on the wall by $D$, then
\begin{equation}
 e^\nu D_\nu e^\mu=\kappa(s)u^\mu,
 \qquad
 \kappa(s)=-a(s)\sinh\chi(s).
 \label{eq:kappa-GC}
\end{equation}
Thus $a(s)$ always denotes the acceleration of the curve in the static BTZ
slice, while $\kappa(s)$ denotes its intrinsic acceleration in the wall.

Let $s= s_*$ be a zero of $a(s)$ for the initial curve. In a $\delta$-neighborhood of $s_*$ choose $\chi(s)$ to satisfy\footnote{The reason for choosing this equation will become clear shortly.} 
\be\label{eq:defchi}
\chi'{:= \frac{d\chi}{ds}}=1-a\cosh\chi
\ee
with $\chi(s_*) = 0$. In a larger $2\delta$-neighborhood of $s_*$ choose $\chi$ to smoothly go to zero. With this choice, $K_{01}(s_*) = -1$. The flat-space Gauss constraint \eqref{eq:Gauss-flat-GC} is then satisfied there simply by taking $K_{00}(s_*)=2$. In the transition
regions, where $a\neq0$, we set
\begin{equation}
 K_{00}=\frac{1-\chi'^2}
 {a\cosh\chi}.
 \label{eq:K00-transition-GC}
\end{equation}
Outside the support of $\chi$ it reduces to the time-symmetric expression $K_{00}=1/a$. Consequently the initial data differ from time-symmetric data only where this is required to pass through the zeros of $a$. The support should be chosen small but finite: taking $\delta$ arbitrarily small forces rapid variation in the transition region and can make $K_{00}$ unnecessarily large.

A flat two-dimensional surface can be either a cylinder or a cone. The simplest way to ensure that it is not a cone is to require that the initial data is invariant under the discrete symmetry $(\tau,s)\rightarrow (-\tau,-s)$ If our initial curve at $\tau=0$ is reflection symmetric, $X^0(-s) = X^0(s)$, this will be satisfied by choosing the boost $\chi(\sigma)$ to be antisymmetric: $\chi(-s) = - \chi(s)$. 

Now that we have constructed our initial data, we can in fact show that there exists a flat timelike cylinder that passes through it (at least for a finite time). Actually, it suffices to solve the Gauss-Codazzi equations, since the fundamental theorem of surface theory ensures that if these equations are satisfied, a surface exists with the given curvature and extrinsic curvature \cite{Baer,Leon-Guzman}.\footnote{This theorem directly applies to the $AdS_3$ covering space of BTZ, but since the Gauss-Codazzi solution will be periodic, the resulting surface will be periodic and descends to a surface in BTZ.}

Choose $\tau$ so that the curves of constant $s$ are timelike geodesics of the induced metric and are orthogonal to $C$ at $\tau=0$. Since the desired induced metric is flat, Gaussian normal coordinates give
\begin{equation}
 q=-d\tau^2+h(\tau,s)^2ds^2,
 \qquad
 h(\tau,s)=1+\tau\kappa(s),
 \label{eq:flat-Fermi-GC}
\end{equation}
for sufficiently small $|\tau|$. The result \eqref{eq:flat-Fermi-GC} follows from the observations that $h(0,s)=1$,
$\partial_\tau h(0,s)=\kappa(s)$, and that flatness is equivalent to
$\partial_\tau^2h=0$. As usual, such coordinates remain valid until the geodesics cross and form caustics.

Let
\begin{equation}
 u=\partial_\tau,
 \qquad
 e=h^{-1}\partial_s,
 \qquad
 H(\tau,s)=\frac{\kappa(s)}{h(\tau,s)}.
 \label{eq:Fermi-frame-GC}
\end{equation}
Let $K_{00}$, $K_{01}$, and $K_{11}$ be defined as in
\eqref{eq:K-components-GC}. The Gauss-Codazzi equations are
\begin{align}
 K_{01}^2-K_{00}K_{11}&=1,
 \label{eq:Gauss-evolve-GC}\\
 \partial_\tau K_{01}
 &=\frac{1}{h}\partial_s K_{00}-2H K_{01},
 \label{eq:Codazzi1-GC}\\
 \partial_\tau K_{11}
 &=\frac{1}{h}\partial_s K_{01}-H(K_{00}+K_{11}).
 \label{eq:Codazzi2-GC}
\end{align}
The initial values at $\tau=0$ are those constructed above:
\begin{equation}
 K_{11}=-a\cosh\chi,
 \qquad
 K_{01}=-\chi',
 \qquad
 K_{00}=\frac{1-\chi'^2}{a\cosh\chi},
 \label{eq:initial-K-all-GC}
\end{equation}
Away from $K_{11}=0$, one may use \eqref{eq:Gauss-evolve-GC} algebraically to eliminate
$K_{00}$. The resulting two equations for $K_{01},K_{11}$ form a strictly hyperbolic first-order system so smooth initial data have a unique smooth solution for a sufficiently small finite time.
Furthermore, the equations that define $K_{\mu \nu}$ can be integrated to show that we may indeed take the associated surface in BTZ to contain our curve $C$ and to induce the specified initial data on $C$. This completes the argument for finite-time existence of our surface away from points where $a=0$.

A solution in a neighborhood of each zero of $a$ can be constructed explicitly. Define the future-directed null vector
\be 
L^\mu = u^\mu + e^\mu
\ee
where $u^\mu$ and $e^\mu$ are given by \eqref{eq:boosted-frame-GC} and \eqref{eq:em-GC}. We now shoot affinely parametrized null geodesics from $C$ with initial tangent $L^\mu$. It is shown in Appendix \ref{app:null-geodesic-wall} that the induced metric on the resulting surface will be flat provided \eqref{eq:defchi} is satisfied. This is the reason why that differential equation was chosen. Furthermore, since the unit normal on the initial curve is still $n^\mu$, this flat surface has the same extrinsic curvature on the $T=0$ surface as our initial data. So away from the zero of $a$, this solution must agree with the one obtained from the hyperbolic equations.

To see that the above solution can describe a surface that enters the horizon, we can start with a curve that is arbitrarily close to the horizon at one point, such as 
\be
X^0(\sigma) = A(1 + \cos\sigma) + \epsilon
\ee
where $0<A<1$. The lower limit on $A$ comes from the fact that a constant radius circle stays at that radius and never enters the horizon, while the upper limit comes from the fact that at $T=0$, the Kruskal coordinate $X$ is restricted to $|X| <2$. If $\epsilon = 0$, our initial curve is in fact tangent to the horizon at one point. Under any evolution in $\tau$ it will then enter the horizon near that point. By continuity, for small $\epsilon \ne 0$, the surface will still enter the horizon even though it starts strictly outside. 

{Note also that since our construction above applied to general curves $C$, this particular $C$ can be obtained by smoothly deforming initial data that gives the static wall at $X=constant$; i.e., it can be obtained by smoothly deforming the usual boxed black hole solution. Furthermore, at each step in the deformation the initial data defines a corresponding finite-time solution.}

Once a piece of the wall enters the horizon, the entire wall must enter as well. This follows since the intersection of our timelike wall with any null surface (such as the event horizon) must be a spacelike or null curve which wraps around the wall in a finite amount of time. Said another way, once a point on the wall enters the horizon, the entire causal future of that point must lie inside the black hole. But on the wall, the causal future of any point includes an entire $S^1$ cross-section. We are assuming here that evolution continues until the surface is inside the black hole. If it should end outside the black hole, that would be a more dramatic ``end of time" for the Dirichlet wall problem.

Finally, let us close this section with a few comments about the energy of our solutions. Since $X=0$ is a minimal surface in the time-symmetric $T=0$ surface, its extrinsic curvature vanishes in this surface. For $\epsilon=0$ our wall is tangent to $X=0$ at just one point, bending outward from that surface. It must therefore be concave there rather than convex. This means that the $tt$ component of the Brown-York stress tensor has the same sign as in the AdS cosmologies studied in section \ref{sec:examples}; i.e., as advertised our solution is locally like a wall with $E>0$.

On the other hand, we can easily arrange for the total energy to be negative. To do so, first note that the contribution of the above near-horizon region of negative Brown-York energy density will be heavily suppressed by the large redshift near the horizon. Note also that any region of the wall over which the angular coordinate $\phi$ is constant (i.e., independent of both $X$ and $T$) would have vanishing Brown-York energy density due to the reflection symmetry across the wall. While our ansatz above does not allow regions in which $\phi$ is exactly constant, it does allow us to come arbitrarily close. At $T=0$ we may thus connect the above near-horizon region of our Dirichlet wall to a region far from the black hole by using nearly-static and nearly-radial segments which again contribute arbitrarily small amounts of energy. The total energy of such domain walls is thus dominated by the part of the wall that lies far from the black hole and which can be taken to form a convex surface. The dominant region thus has negative Brown-York energy density, and the total energy is negative as desired.

\section{Extensions to higher dimensions}
\label{sec:trapped}

It is straightforward to extend to higher dimensions the construction of initial walls analogous to that studied in Sec. \ref{sec:WallinBTZ}. However, a significant difference then arises from the fact that higher dimensional Einstein-Hilbert gravity {\it does} have local degrees of freedom. The absence of such degrees of freedom in the 2+1 context is what allowed us to conclude that any solution must be given by finding a surface with the given induced metric in the original BTZ spacetime, and that there could be no additional back-reaction of the Dirichlet wall on the bulk metric. 

The argument clearly fails in higher dimensions, where the wall can be expected to emit and absorb gravitational radiation. Indeed, if there were no such emission or absorption then with generic initial data it would be impossible to construct even a power series solution\footnote{This follows from a simple counting argument. A codimension-1 surface can be specified by one function. But, even taking into account the freedom to choose new coordinates on the wall, this is not enough freedom to set all components of the wall-metric to their desired values in a generic spacetime (even if the desired metric tensor is present on the wall at some initial time). }. 

Allowing the bulk to evolve, we can again construct a power series solution as follows. Beginning with the corresponding $T=0$ slice of e.g. AdS-Schwarzschild, one can choose a surface in this slice, and one can set the initial velocity of the surface to vanish everywhere (i.e., one can take the infinitesimal surface near $T=0$ to share the $T\rightarrow -T$ symmetry of the background spacetime). Expanding both the bulk metric and the location of the wall in powers of $T$ then allows one to construct a power series solution. One may then correspondingly argue that one can again arrange part of the wall to pass through the surface that, in a pure AdS-Schwarzschild solution, would be called the black hole event horizon.

While features of finite-time evolution are then difficult to predict, what the above power series solution does show is that the wall has entered the causal future of a trapped surface. In the context of the Penrose singularity theorem (and in particular without a finite-distance boundary), this would imply the formation of a singularity in the future, as well as a corresponding event horizon. Whether similar behavior occurs in the presence of the wall is an interesting question whose future study we encourage.

However, we wish to point out that there is another context in which one can find a trapped surface that includes part of the Dirichlet wall, and which does not in fact require the initial presence of a black hole. In particular, this is the case when two high-energy particles collide near the Dirichlet wall.

In a linearized approximation, the presence of a Dirichlet wall with flat induced metric causes each particle to see an image particle with negative mass. The image particle exerts a repulsive force on the original particle, so particles are repelled from the Dirichlet wall. This raises the question of whether uncharged\footnote{Black holes with extremal charge experience no such effective potential at the wall since there is also a counter-balancing electrostatic attraction from an image electric charge if we also impose Dirichlet boundary conditions on the Maxwell field. In fact, \cite{Andrade:2015qea} found that an effective kinetic term of such extremal black holes becomes negative near the wall, signalling another potential instability.} black holes can form arbitrarily close to a Dirichlet wall in the full nonlinear theory. We now argue that at least trapped surfaces can form near the wall though, as above, whether that leads to an event horizon remains a topic for future investigation. 

 We study the collision of two high energy particles moving parallel to a flat Dirichlet wall. In the absence of the wall, each particle can be described by an Aichelburg-Sexl shock wave\footnote{Strictly speaking, this describes the particle in the limit as their rest mass goes to zero and their velocity approaches the speed of light, keeping their total energy, $\mu$, fixed.}, and it has been shown that a marginally trapped surface can form in the collision of two such shock waves. 
 For simplicity, we only consider head-on collisions. We will show that a Dirichlet wall (parallel to the colliding particles) has no effect on the marginally trapped surface that forms provided that the center of mass energy, $2\mu$, is small enough that this surface does not hit the wall. We will also discuss what happens at higher energies. We will work in four dimensions, but similar results hold for $D>4$.

The Aichelburg-Sexl metric for a single particle moving in the $\bar z$ direction at $\bar x = L$ and $\bar y = 0$ is 
\begin{equation}\label{eq:AS}
ds^2
= - d\bar ud \bar v + d \bar x d \bar x+ d \bar y d \bar y+ \Phi(\bar x, \bar y)\delta(\bar u)d\bar u^2
\end{equation}
where $\bar u = \bar t - \bar z$, $\bar v = \bar t + \bar z$, and $\Phi$ satisfies Poisson's equation in the flat $\bar x, \bar y$ coordinates:
\begin{equation}
\nabla^2 \Phi = -16 \pi \mu \delta(\bar x -L)\delta(\bar y)
\end{equation}
Since $\Phi$ satisfies a linear equation, we can add a Dirichlet wall at $\bar x = 0$ with flat induced metric by adding an image particle with negative mass moving along a parallel worldline at $\bar x = -L$. The potential $\Phi$ will now satisfy
\begin{equation}\label{eq:Dphi}
\nabla^2 \Phi = 16 \pi \mu [\delta(\bar x +L) - \delta(\bar x -L)]\delta(\bar y),
\end{equation}
with solution 
\be\label{eq:Phi}
\Phi(\bar x,\bar y) = 4\mu \log \frac{(\bar x + L)^2 + \bar y^2}{(\bar x - L)^2 + \bar y^2}.
\ee

Clearly an identical particle moving in the $-\bar z$ direction will be described by the metric \eqref{eq:AS} with $\bar u$ and $\bar v$ interchanged. By causality, it can have no influence on the first particle until they collide. So we can obtain an exact metric everywhere outside the future of the collision by simply superposing these two solutions.

Since the above metric contains $\delta$-functions, geodesics and their tangent vectors will appear discontinuous across the shock. It is thus better to work with the following coordinates in which the metric is continuous:
\be
\begin{aligned}
    \bar u & =u, \\
    \bar v & = v + \Phi \Theta(u) + \frac{1}{4}u\Theta(u) (\nabla \Phi)^2,\\
    \bar x^i & = x^i +\frac{u}{2} \nabla_i \Phi \ \Theta(u),
\end{aligned}
\ee
where $\Theta(u) $ is the step function and we have combined $x,y$ into $x^i$. Note that $\Phi(x^i)$ is the same function as in \eqref{eq:Phi} but with $\bar x^i$ replaced by $x^i$. In these new coordinates, the metric still takes the simple flat space form \eqref{eq:AS} (without the bars) for $u\le 0$ and $v\le 0$. However just past the $u=0$ shock wave, i.e. for $v < 0$ and small $u> 0$, the metric becomes
\be
ds^2 = -du dv + [\delta_{ij} + u \nabla_i \nabla_j \Phi + O(u^2) ] dx^i dx^j,
\ee
and similarly for the region just past the $v=0$ shock.

To find the apparent horizon in the collision of the shock waves, we follow the procedure used by Eardley and Giddings \cite{Eardley:2002re}. We will construct the surface by gluing together two disks along their edge. One disk lives in the $u=0$ null surface and the other lives in the $v=0$ null surface. We start with the $u=0$ disk defined by $v = -\Psi(x^i)$, where $\Psi$ is a positive function that vanishes on some curve ${\cal C}$. The other half is an identical surface on $v=0$ given by $u = -\Psi(x^i)$. These two disks clearly meet at $u = v = 0$ along ${\cal C}$. Since we want the orthogonal null vectors to be continuous at ${\cal C}$ we require that 
\be\label{eq:cont}
\nabla_i \Psi \nabla^i \Psi =4 \quad {\rm on} \quad {\cal C}.
\ee
This follows from the fact that the outward null normal, $\ell$, to both $u = 0$ and $v+\Psi(x^i) = 0 $ is 
\be
-\ell = dv + \nabla_i\Psi dx^i + \frac{1}{4} (\nabla\Psi)^2 du.
\ee
The null normal to $v= 0 $ and $u+\Psi = 0$ is similar with $u$ and $v$ interchanged. So they agree when \eqref{eq:cont} is satisfied.

The divergence of $\ell$ vanishes if
\be 
\nabla^2 (\Psi - \Phi) =0.
\ee
So $\Psi$ depends on $\Phi$ only through its Laplacian, which by \eqref{eq:Dphi} is just the sum of two delta-functions. To get a nonzero solution for $\Psi$, we clearly want the curve ${\cal C}$ to enclose the delta-function at $\bar x = L$. But since the spacetime ends on the Dirichlet wall at $\bar x = 0$, $\Psi$ does not know about the second delta-function. The solution is thus the same as if there was no wall:
\be
\Psi(x,y) = -4\mu \log [(x-L)^2 + y^2] + c_0,
\ee
where $c_0$ is a constant. Imposing \eqref{eq:cont}, we find that ${\cal C}$ must be the curve 
\be
(x-L)^2 + y^2 = 16\mu^2.
\ee
Choosing the constant $c_0$ so that $\Psi$ vanishes on ${\cal C}$ yields
\be
\Psi(x,y) = -4\mu \log \frac{(x-L)^2 + y^2}{16\mu^2}. 
\ee
This is the same result that Penrose first found \cite{Penrose} (with $L=0$) for the apparent horizon in the collision of two high energy particles. So even though the presence of the Dirichlet wall changes the metric \eqref{eq:AS} via \eqref{eq:Phi}, it does not change the location of the marginally trapped surface.

If the total energy of each particle satisfies $\mu < L/4$, then the marginally trapped surface we have found does not reach the Dirichlet wall. {However, if the particle satisfies $\mu = L/4$ exactly, the surface will intersect the Dirichlet wall at a single point (at $u = v = x = y = 0$.) The boundary point is still contained in the manifold, and has a tangent space which is a four-dimensional vector space, as usual.\footnote{The tangent space is defined as the vector space generated by derivations of smooth functions on the manifold at the point $p$.} We may thus define the future-directed null normal vectors as usual, and find that their expansion is again nonpositive. The surface is smooth by construction, and thus we have found a marginally trapped surface which intersects the Dirichlet wall.}

This argument does not show that the marginally trapped surface is an apparent horizon (which must be the outermost marginally trapped surface). But it strongly suggests trapped surfaces will form. In asymptotically flat spacetimes, under reasonable assumptions one can show that a trapped surface cannot lie in the past of future null infinity. It must be inside a black hole. We do not know the evolution to the future of the collision with the Dirichlet wall, but it is plausible that it will enter the horizon of a black hole. If it does, evolution must stop as in the example in the previous section.

\section{Discussion}
\label{sec:disc}
Our work above explored solutions to Einstein gravity with Dirichlet walls and found that they often end in finite time. 

We initially focused on spacetimes formed by inserting $S^{d-2} \times {\mathbb R}$ Dirichlet walls into vacuum AdS$_d$-cosmologies. We argued that for all $d\ge 3$ we can find spacetimes that form singularities that are everywhere spacelike and at which the volume of space shrinks to zero. Since they cut off the evolution everywhere in space, we called these end-of-time singularities. Detailed calculations were provided for $d=3$, in which case we also constructed a generalization by inserting the above Dirichlet walls into 
AdS$_3$-cosmologies with conical defects. For the one-parameter family studied in section \ref{sec:conicalsingularities}, the insertion of the wall in fact excised the part of the 
AdS$_d$-cosmology containing the defect, so that no conical singularity remained in the physical space. A further conical singularity in the physical space was then considered in appendix \ref{sec:pointmass}. Such end-of-time singularities arise under finite time-evolution despite the fact that the local-in-time well-posedness theorem of \cite{anWellposedGeometricBoundary2025a} applies to all of these spacetimes (see footnote \ref{foot:AA}).

The above solutions have the opposite sign ($E>0$) of the conserved energy from that of familiar `boxed black hole' solutions describing a static Dirichlet wall surrounding a black hole (which have $E<0$). However, we argued that this difference is not fundamental as in various contexts one can drive the system from one regime to the other by imposing time-dependent boundary conditions on the wall. 

While any solution where $E$ changes sign will violate the assumptions of \cite{anWellposedGeometricBoundary2025a} at some time, we argued that it is physically important to allow such failures and, in particular, that the conditions of \cite{anWellposedGeometricBoundary2025a} are generally not preserved under finite time evolution in the presence of gravitational radiation. This leaves one with the choice of simply ending the time evolution when the condition fails, regardless of whether the fields might still be smooth, or of considering spacetimes that violate the condition. We have chosen the latter here.

We also argued that Dirichlet walls can encounter singularities even when $E<0$. In particular, we constructed solutions in which part of an $S^1\times {\mathbb R}$ Dirichlet wall falls across the horizon of a BTZ black hole. It then {followed} that the entire wall must fall in. While we can have $E<0$ for such solutions\footnote{In particular, $E$ for our solutions is always real. This is in sharp contrast to the complex energies that one obtains by considering static (rotationally-symmetric) boxed black hole solutions and analytically continuing the radius $r_0$ of the wall to values smaller than the radius of the black hole's event horizon (as arises e.g. in discussions of $T\bar T$ deformations \cite{mcgoughMovingCFTBulk2018}.}, the $tt$ component of the Brown-York stress tensor is locally positive in our solutions where the wall first crosses the horizon. In AdS$_3$ this guarantees that the wall will encounter a singularity. In higher dimensions, we can argue in several contexts that the wall touches a marginally trapped surface, but whether this then guarantees the formation of singularities remains to be investigated in more detail. In general, our work complements studies of stability of gravity in the more familiar boxed black hole context with various boundary conditions on a finite-distance wall \cite{Andrade:2015gja,Anninos:2024xhc,liuNewWellPosedBoundary2024,Anninos:2024wpy}.

It might also be interesting to consider other (non-Dirichlet) boundary conditions on a finite-distance wall, such as Neumann or mixed boundary conditions. (Conformal boundary conditions are not well-posed \cite{Liu:2025xij}.) We expect similar solutions in these cases also. If one fixes a general linear combination of the metric, extrinsic curvature, and higher derivatives thereof, in the long-wavelength limit this will again reduce to a Dirichlet boundary condition. We thus expect similar results in that limit for generic such boundary conditions.

In addition to the above classical open questions, one would like to explore the effect of various possible quantum and/or stringy corrections. It is also of interest to understand how our singularities might be described in any field-theory-like dual, and in particular what they might imply for $T\bar T$ deformations of AdS$_3$/CFT$_2$ or for analogues in other dimensions.

\section*{Acknowledgements}
We are grateful for interesting discussions with Tom Banks, Damian Galante, Eva Silverstein, and with the participants of the December 2025 Simons Center for Geometry and Physics workshop {\it Timelike Boundaries in Classical and Quantum Gravity}. 
This work was supported in part by NSF grant PHY-2408110, and by funds from the University of California.

\appendix
\section{Adding point masses}
\label{sec:pointmass}

Having constructed a set of vacuum solutions in Sec. 2, we may also include point masses in the spacetime. As usual in $d = 3$, we model such a mass by including a conical singularity along a geodesic~\cite{Gott:1982qg}. If we want to keep our Dirichlet wall centered at the origin, we can introduce the conical singularity to the vertices of the fundamental domain, in order to introduce a singularity inside of the spacetime.

Keeping the singularity at the origin (labeled $\delta_0$) and introducing a new conical singularity at the vertices (labeled $\delta_m$), the area of the fundamental domain becomes $A = 4 \pi (g-1) + \delta _0 + \delta_m.$ The parameter $\alpha$ is thus
\begin{equation} \label{eq:alphadeficitvertices}
    \alpha = \frac{\cot \left(\frac{2 \pi - \delta_m}{8 g}\right)}{\sqrt{\cos \left(\frac{\delta _0-\delta _m}{8 g}\right) \cos \left(\frac{4 \pi - (\delta_0 + \delta_m)}{8 g}\right) \csc ^2\left(\frac{2 \pi - \delta _m}{8 g}\right)}}.
\end{equation}

After inserting the Dirichlet boundary centered at the origin, we can again find the singularities. Now, the singularities first form at
\begin{equation}
\begin{split}
    \tau_i &= \cos ^{-1}\left(\frac{L \sqrt{\alpha ^2-1} }{2 \pi - \delta_0}\right) \\
    &= \cos ^{-1}\left(\frac{L}{2 \pi -\delta _0} \sqrt{\cos ^2\left(\frac{2 \pi - \delta _m}{8 g}\right) \sec \left(\frac{\delta _0-\delta _m}{8 g}\right) \sec \left(\frac{4 \pi - (\delta _0 + \delta_m)}{8 g}\right)-1}\right),
\end{split}
\end{equation}
which, in terms of $\alpha$, is identical to Equation~\eqref{eq:initialtimedeficit}. The point mass only enters in the value of $\alpha$. 

The entire boundary becomes singular at 
\begin{equation}
\begin{split}
    \tau_f &= \cos ^{-1}\left(\frac{L}{2 \pi -\delta _0} \sqrt{\alpha ^2 \cos ^2\left(\frac{2 \pi - \delta _0}{8 g}\right)-1}\right) \\
    &= \cos ^{-1}\left(\frac{L}{2 \pi - \delta _0} \sqrt{\cos ^2\left(\frac{2 \pi - \delta _0}{8 g}\right) \cos^2\left(\frac{2 \pi - \delta _m}{8 g}\right) \sec \left(\frac{\delta _0-\delta _m}{8 g}\right) \sec \left(\frac{4 \pi - (\delta _0+\delta _m)}{8 g}\right)-1}\right),
\end{split}
\end{equation}
which is also identical to Equation~\eqref{eq:finaltimedeficit} as a function of $\alpha$. 

In this case, the singularities again lie at
\begin{equation}
    \left(\tau ,~\sinh ^{-1}\left(\frac{L \sec (\tau )}{2 \pi -\delta_0 }\right),~ \frac{n \pi}{2 g} \pm \cos ^{-1}\left(\frac{(2 \pi -\delta_0 )}{\alpha L} \sqrt{\frac{L^2}{(2 \pi - \delta_0)^2}+\cos ^2(\tau )}\right)\right),
\end{equation}
which is again identical in form to Equation~\eqref{eq:singularitydeficit} as a function of $\alpha$. The effect of the point mass arises in $\alpha$ only.

For this solution, the maximum bound on the size of the Dirichlet boundary, enforced by requiring that no singularities occur in the $\tau = 0$ plane, is again identical to Eq.~\eqref{eq:Lbound} as a function of $\alpha$. To ensure that the singularity is everywhere spacelike, the boundary circumference must satisfy 
\begin{equation}
\begin{split}
    L &< \frac{\left(2 \pi -\delta _0\right)}{\sqrt{\alpha ^2-1}} \sqrt{1-\alpha ^2 \sin ^2\left(\frac{2 \pi- \delta _0}{8 g}\right)} \\
    & \quad = \frac{\left(2 \pi -\delta _0\right)}{2 \sqrt{2}} \sqrt{\csc^2\left(\frac{2 \pi - \delta _0}{8 g}\right)} \\
    & \qquad \times
    \sqrt{6 \cos \left(\frac{2 \pi - \delta _0}{4 g}\right)+\cos \left(\frac{\delta _0-\delta _m}{4 g}\right)+2 \cos \left(\frac{2 \pi - \delta _m}{4 g}\right)+\cos \left(\frac{4 \pi - (\delta _0+\delta _m)}{4 g}\right)-2},
\end{split}
\end{equation}
which is again identical to Eq.~\eqref{eq:spacelikeupperbound2}.

Finally, the extrinsic curvature, quasilocal stress-energy tensor, and conserved energy can be computed. The results are identical to the vacuum case. The conserved energy is independent of $\delta_m$. This is expected since adding the point mass does not change the solution near the Dirichlet wall.

\section{Null-geodesic construction of the flat BTZ wall }
\label{app:null-geodesic-wall}

In this appendix we show that the induced metric on the surface generated by the family of null geodesics in Sec. 3 is flat. As in that section, we set the AdS radius to one, so the BTZ black hole has $R_{\mu\nu}=-2g_{\mu\nu}$.

Let $C$ be the initial spacelike curve, parametrized by proper length $s$, with
unit tangent $e^\mu$. Using the notation of Sec.~3, let
\begin{equation}
 u^\mu=\cosh\chi\,t^\mu+\sinh\chi\,m^\mu,
 \qquad
 n^\mu=\sinh\chi\,t^\mu+\cosh\chi\,m^\mu,
 \label{eq:app-boosted-frame}
\end{equation}
and define the future-directed null vector
\begin{equation}
 L^\mu=u^\mu+e^\mu .
 \label{eq:app-null-L}
\end{equation}
Define the acceleration $a(s)$ by $e^\nu\nabla_\nu e^\mu=a(s)m^\mu$. Then the fact that the static BTZ slice
is totally geodesic gives
\begin{equation}
 e^\nu\nabla_\nu L^\mu
 =\kappa(s)L^\mu+B(s)n^\mu,
 \qquad
 \kappa=-a\sinh\chi,
 \qquad
 B=\chi'+a\cosh\chi,
 \label{eq:app-L-derivative}
\end{equation}
where a prime denotes $d/ds$.

We now shoot affinely parametrized null geodesics from $C$ with initial tangent
$L^\mu$. Thus $X^\mu(s,v)$ satisfies
\begin{equation}
 \frac{\partial^2X^\mu}{\partial v^2}
 +\Gamma^\mu_{\alpha\beta}(X)
 \frac{\partial X^\alpha}{\partial v}
 \frac{\partial X^\beta}{\partial v}=0,
 \qquad
 X^\mu(s,0)=C^\mu(s),
 \qquad
 \left.\frac{\partial X^\mu}{\partial v}\right|_{v=0}=L^\mu(s).
 \label{eq:app-null-geodesic}
\end{equation}
Set
\begin{equation}
 k^\mu=\partial_vX^\mu,
 \qquad
 J^\mu=\partial_sX^\mu ,
 \label{eq:app-kJ}
\end{equation}
 so these two vectors form a basis of tangent vectors to our surface. 
Then $k^\nu\nabla_\nu k^\mu=0$, $k_\mu k^\mu=0$, and, since
$[\partial_v,\partial_s]=0$,
\begin{equation}
k^\nu \nabla_\nu J^\mu=J^\nu\nabla_\nu k^\mu .
 \label{eq:app-commuting-variation}
\end{equation}
It follows immediately that $k_\mu J^\mu$ is independent of $v$. Since
$L_\mu e^\mu=1$ on $C$, $k_\mu J^\mu=1$ so the induced metric on the surface, $q_{ij}$, has components
\begin{equation}
 q_{vv}=0,
 \qquad
 q_{sv}=1.
 \label{eq:app-first-metric-components}
\end{equation}

To find the remaining component, $q_{ss}$, we use the equation of geodesic deviation 
\begin{equation}
 k^\gamma\nabla_\gamma(k^\beta\nabla_\beta J^\mu)
 =R_{\alpha\beta\gamma}{}^\mu J^\alpha
   k^\beta k^\gamma
 =-k^\mu,
 \label{eq:app-jacobi}
\end{equation}
where in the last equality we used 
the fact that BTZ has constant curvature. Defining
$W^\mu=k^\beta \nabla_\beta J^\mu$, we also have $k\cdot W=0$, so
\begin{equation}
 \partial_v(W_\mu W^\mu)=2W_\mu k^\gamma\nabla_\gamma W^\mu=-2W_\mu k^\mu=0.
 \label{eq:app-Wnorm}
\end{equation}
At $v=0$, Eq.~\eqref{eq:app-L-derivative} gives
$W^\mu=\kappa L^\mu-B n^\mu$, and therefore
\begin{equation}
 W_\mu W^\mu=B(s)^2
 \label{eq:app-Wnorm-B}
\end{equation}
along the entire null geodesic.

It remains only to determine $q_{ss}=J_\mu J^\mu$. From
Eqs.~\eqref{eq:app-jacobi} and \eqref{eq:app-Wnorm-B},
\begin{equation}
 \partial_v^2(J_\mu J^\mu)
 =2W_\mu W^\mu+2J_\mu k^\gamma\nabla_\gamma W^\mu
 =2\bigl(B^2-1\bigr).
 \label{eq:app-qss-second}
\end{equation}
The initial conditions are
\begin{equation}
 J_\mu J^\mu\big|_{v=0}=1,
 \qquad
 \left.\partial_v(J_\mu J^\mu)\right|_{v=0}=2\kappa,
 \label{eq:app-qss-initial}
\end{equation}
so the induced metric is exactly
\begin{equation}
 q=\left[1+2v\kappa(s)+v^2\bigl(B(s)^2-1\bigr)\right]ds^2
   +2\,ds\,dv .
 \label{eq:app-induced-general}
\end{equation}
There are no higher powers of $v$. For a metric of the form
$q=F(s,v)ds^2+2ds\,dv$, the scalar curvature is
\begin{equation}
 {\cal R}[q]=\partial_v^2F.
 \label{eq:app-R-F}
\end{equation}
Thus the wall is flat precisely when $B^2=1$. Choosing the branch used in
Sec.~3 gives the regular first-order equation
\begin{equation}
 {\chi'=1-a\cosh\chi.}
 \label{eq:app-chi-ode}
\end{equation}
For this choice,
\begin{equation}
{
q=\bigl(1+2v\kappa(s)\bigr)ds^2+2\,ds\,dv,
 \qquad
 \kappa=-a\sinh\chi ,
 }
 \label{eq:app-induced-flat}
\end{equation}
and $\det q=-1$.

One cannot use this null geodesic construction everywhere on our initial curve to obtain a flat cylinder since the solution to \eqref{eq:app-chi-ode} will typically not have the right periodicity around the circle. However at a point where the acceleration of the
initial curve vanishes, $a(s_*)=0$, Eq.~\eqref{eq:app-chi-ode} simply gives
$\chi'(s_*)=1$; neither the construction nor the induced metric becomes
singular there.

\bibliographystyle{JHEP}

\bibliography{references}

\end{document}